\documentclass[preprint,showpacs,preprintnumbers,amsmath,amssymb]{revtex4}
\usepackage{graphicx}
\usepackage{bm}

\usepackage{epstopdf}
\begin{document}
\title{Two dimensional vortex structures with multi scale interactions in thin film superconductors}

\author{Yu. D. Fomin, V. N. Ryzhov, E. N. Tsiok}
\affiliation{Institute for High Pressure Physics RAS, Kaluzhskoe
shosse, 14, Troitsk, Moscow, 108840 Russia}

\date{\today}

\begin{abstract}
Interaction of particles of many systems can be effectively
approximated by multiscale interaction potentials. Such potentials
are widely used for investigation of colloidal systems and
colloid-polymer mixtures, complex liquids (for instance, water)
and other system. Also, it is of particular interest that
effective interaction of vortices in a thin superconducting film
can be approximated by a multiscale potential. The subject matter
of the present paper is a system like this. Based on a multiscale
potential we have shown that this system demonstrated a large
number of different phases. It is particularly important that we
analyze the influence of a long-range interaction of the system
properties and show that properly taking into account long-range
forces results in a dramatic change in the system phase diagram.
\end{abstract}


\maketitle

\section{Introduction}

Two dimensional (2d) systems are of great interest for both
fundamental science and technological applications. In many
aspects 2d systems are basically different from three dimensional
(3d) ones. First of all 2d melting is a much more complex
phenomenon than the melting of 3d crystals. While 3d crystals
always melt via the first order phase transition, 2d systems can
melt via at least three different scenarios \cite{ufn}.

Another question of interest in 2d systems is their crystalline
structure. Until recently, only a simple triangular crystal which
is a close packed structure in 2d was known experimentally. The
situation got a lot of attention when graphene was discovered
\cite{graphene}. This discovery gave impetus to further
experimental investigation of possible ordered structures of 2d
systems and some of them were really discovered. For example, in
Ref. \cite{geim} a square phase of water (square ice) was observed
when water was strongly confined between two graphene sheets. The
square crystal was also observed in one atom thick film of iron
confined between graphene sheets \cite{iron}.

Another example of systems where complex 2d structures can be
found is 2d colloidal suspensions. For example, in Ref.
\cite{dobnikar} a system of colloidal particles in a magnetic
field was studied. The authors changed the system density
intending to find the same sequence of phases which was obtained
in their model system from Refs. \cite{camp1,camp2} and clearly
observed a square phase in their colloidal system.

An important class of 2d systems is the systems of vortices in
superconducting films predicted by Abrikosov in 1957
\cite{abrikosov} and experimentally observed for the first time by
Essmann and Trauble in 1967 \cite{abrikosov-exp}. These important
results gave rise to a novel field in the studies of
superconductivity. Numerous theoretical and experimental works are
devoted to the lattices of vortices in superconductors. Like
atomic and colloidal particles, the vortices usually form a
triangular lattice. However, there are some experimental works
where other types of ordered structures were observed ( see, for
instance, Ref. \cite{chains} for chains of vortices and
\cite{vortices-sq} for the square phase of vortices).

One can see that attempts to find non-triangular one-component 2d
structures have been made in different classes of systems.
However, the progress is not very fast. At the same time a
theoretical search for novel structures in 2d systems is very
active and numerous different structures have been found. Most of
the theoretical studies are concerned with so called core-softened
systems, i.e.  systems with the negative curvature of the
potential. Such systems can serve to describe effective
interactions in many real systems, such as colloidal systems and
colloid-polymer mixtures, water, liquid metals, some molecular
liquids, etc. This potential is characterized by the presence of
two length scales. The competition between these length scales
leads to complex structure formation in the system. In the above
mentioned publications Refs. \cite{camp1,camp2} exactly the
core-softened potential was used which allowed the authors of
those works to find several unusual crystalline structures such as
square crystal and Kagome lattice. A cluster liquid was also
observed. Many other authors studied other core-softened systems
and found the formation of numerous phases (see, for instance,
\cite{malescio,nk,kahl,engel,blanco,we,we2,we3,we4,we5,we6,truskett}
and references therein). It was shown that 2d systems could form
even quasi-crystalline phases (see, for instance,
\cite{engel,we5,wemolphys}). Some works reported particularly
interesting phases \cite{bishop} such as clusters (clump phase in
\cite{bishop}), holes (anti-clump in \cite{bishop}) and
ribbon-like stripes. Such phases as clusters or stripes were also
observed in another system with an interaction similar to the one
which was expected between the vortices in a superconductive film
\cite{njp}.

The interaction potential of vortices in a superconductive film is
usually rather simple \cite{pearl,r1,r2,r3}. However, much more
complex vortex interactions can exist in systems consisting of
layers of different superconductors, in particular, bilayers of
type-I - type-II superconductors, which are widely discussed in
the literature now. For example, in Refs. \cite{xu,xu1} the
interaction consists of long-range logarithmic repulsion and
short-range attraction between the vortices. The authors report
that this system can form different complex phases, such as
circular clusters, ribbons, etc. In several publications
potentials are used which consist of combinations of Bessel
functions \cite{njp,zhao12,zhao17,mendez,struvefunc}. In those
systems such very complex structures as clusters, labyrinths, etc
were also found.

In the present paper we consider the phenomenological interactions
introduced in Refs. \cite{meng1,meng2,meng3}. These interactions
are characterized by core-softened potentials, and therefore one
can expect these systems to demonstrate some complex phases.
Indeed in Refs. \cite{meng1,meng2,meng3} the ground states of
these systems were studied and they were found to form a set of
very unusual structures.

Let us consider the particular system from Refs.
\cite{meng1,meng2,meng3}. In these publications it is assumed that
the interaction potential between the vortices consists of three
parts which are responsible for  intervortex repulsion and for
core overlaps. The particular interaction potential, as suggested
in Refs. \cite{meng1,meng2,meng3} has the following form:

\begin{equation}\label{pot}
  V(r)/V_0=e^{-r/\lambda}-c_2e^{-r/ \xi}+c_3 \lambda
  \frac{tanh[a(r-b)]+1}{r+ \delta}.
\end{equation}
Parameter $\lambda$ sets the unit of length and is chosen to be
unity. Coefficients $c_2$, $c_3$ and $\delta$ are the
phenomenological constants. Parameter $\xi$ is the coherence
length.

Refs. \cite{meng1,meng2,meng3} give very elaborate studies of the
ground state structures of the systems with the potential
(\ref{pot}). In particular, a very interesting phase was obtained
with parameters $c_2=0.2$, $c_3=0.1$, $\delta=0.1$ and $\xi
/\lambda$ from zero up to 10 (see Ref. \cite{meng3}). This phase
consists of big circular clusters. The authors show that in small
systems all particles collapse in a single cluster while if one
increases the number of particles the system splits into several
circular clusters with several hundred of particles each.

This cluster phase looks very unusual and is worth studying in
more detail. In particular, it is interesting to study the
influence of a long-range interaction on this phase. In Refs.
\cite{meng1,meng2,meng3} the potential is considered to extend up
to  half the simulation box. However, the potential from formula
(\ref{pot}) tends to $1/r$ as $r \rightarrow \infty$, i.e. it
becomes Coulomb-like at greater distances. As is known from
studies of coulombic systems, simple cutting of the potential at a
finite distance can only be considered as a first approximation to
the solution \cite{bib}. In the present paper we perform
calculations of system behavior with a simple cut-off of the
interaction potential with more precise treatment of the
long-range interaction and compare the results. We show that the
results basically change if the long-range interactions are taken
into account.

\section{System and methods}

Here we investigate the influence of long-range interactions on
system cluster phase behavior with the potential from (\ref{pot}).
Following Ref. \cite{meng3} we used $c_2=0.2$, $c_3=0.1$,
$\delta=0.1$, $a=2.5$, $b=0.5$ and $\xi /\lambda=4.5$.

In order to take into account the long-range nature of the
potential \ref{pot} we do the following: add and subtract the term
$q^2/r$:

\begin{equation}\label{pot2}
  V_1(r)/V_0=\left( e^{-r/\lambda}-c_2e^{-r/ \xi}+c_3 \lambda
  \frac{tanh[a(r-b)]+1}{r+ \delta} - q^2/r \right) + q^2/r.
\end{equation}
Parameter $q^2=2c_3=0.2$ is selected based on
$q^2/r=lim_{r\rightarrow \infty} V(r)/V_0$.

The term in  brackets is considered as a short-range interaction,
while the last term is calculated by the
Particle-Particle/Particle-Mesh (PPPM or P3M) method, which is
similar in sense but computationally more efficient than Ewald
summation \cite{bib}. The simplest method of performing Ewald
summation in 2d systems is to add slabs of vacuum (see the chapter
'Ewald Summation in a Slab Geometry' of the book \cite{bib}).
 In this method one adds slabs of vacuum at the top and
the bottom of the system and considers it as a three dimensional
one with periodic boundary conditions. The thickness of the vacuum
slab is $10$ from both top and bottom. Importantly, the system
appears to be not "charge neutral". So one should consider it as a
"charged" system in a neutralizing background. The presence of
this background gives an additional term in the system energy:
$E_{neut}=\frac{\pi q_{tot}^2}{2V \alpha ^2}$, where $q_{tot}$ is
the total charge of the system and a is a parameter of Ewald
summation. This term does not depend on the position of the
particles and therefore does not affect the equilibrium structure
of the system.

The short range part of the potential is cut off at distance
$r_c=14$.

Since we add a slab of vacuum the effective particles can leave
the plane and the system will not be two-dimensional any more. In
order to avoid it we introduce two repulsive walls in the system.
The walls are located at height $-0.5$ and $0.5$ and repel the
particles with Lennard-Jones 9-3 potential:

\begin{equation}\label{wall}
 V_{wall}(r)/V_0=\frac{2}{15} \left( \frac{\lambda}{r} \right )^9
 - \left( \frac{\lambda}{r} \right )^3.
\end{equation}

In the present study three types of systems are investigated.
Firstly, we study two-dimensional (2d) systems with the potential
from (\ref{pot}) in order to compare the results with the ones of
Ref. \cite{meng3}. Secondly, we study a quasi two-dimensional
system (q2d) with the same potential. So these systems have slabs
of vacuum and walls at the top and bottom. Finally, we study
systems with a long-range interaction (the potential from
(\ref{pot2})) (coul), slabs of vacuum and walls. At the first step
(from 2d to q2d) we make sure that transition from a
two-dimensional to a quasi two-dimensional system does not affect
its behavior. The second step (from q2d to coul) is necessary to
monitor the influence of  long-range forces on the cluster phase.

We simulated a system of 3200 particles in a box with periodic
boundary conditions. The system had a square shape in the x-y
plane. We performed molecular dynamics simulations. The Lammps
package \cite{lammps} was used. Firstly, we simulated all three
kinds of the systems at temperature $T=0.001$ which is rather low
and the results should be close to the ones at zero temperature.
The densities $\rho=N/A$ where A is the surface of the system
(two-dimensional volume) studied were from $\rho=0.01$ up to
$2.6$. The following densities were considered: $\rho=0.01$,
$0.05$, $0.1$, $0.12$, $0.14$, $0.16$, $0.18$, $0.2$. Then the
densities up to $\rho=2.6$  with step $\Delta \rho =0.2$ were
simulated. After that we also monitored the influence of
temperature by studying the system with $\rho=0.01$ and $\rho=0.2$
up to temperature $T=0.5$. The time step is $dt=0.0001$. Totally,
$5 \cdot 10^6$ steps were performed for system equilibration. In
order to see size effects we repeated the same procedure with a
larger system of $50000$ particles at some densities.

The system structure was analyzed using several methods. We
calculated the radial distribution functions (RDF), order
parameters $\psi_6$ which will be described below and the
distribution of the nearest neighbors in the systems. Two types of
the RDFs were considered: vortex-vortex (vv) and cluster-cluster
(cc). The cc-RDFs were defined as the radial distribution
functions of the centers of mass of clusters.

\bigskip

\subsection{Low densities}

\begin{figure}
\includegraphics[width=18cm,height=22cm]{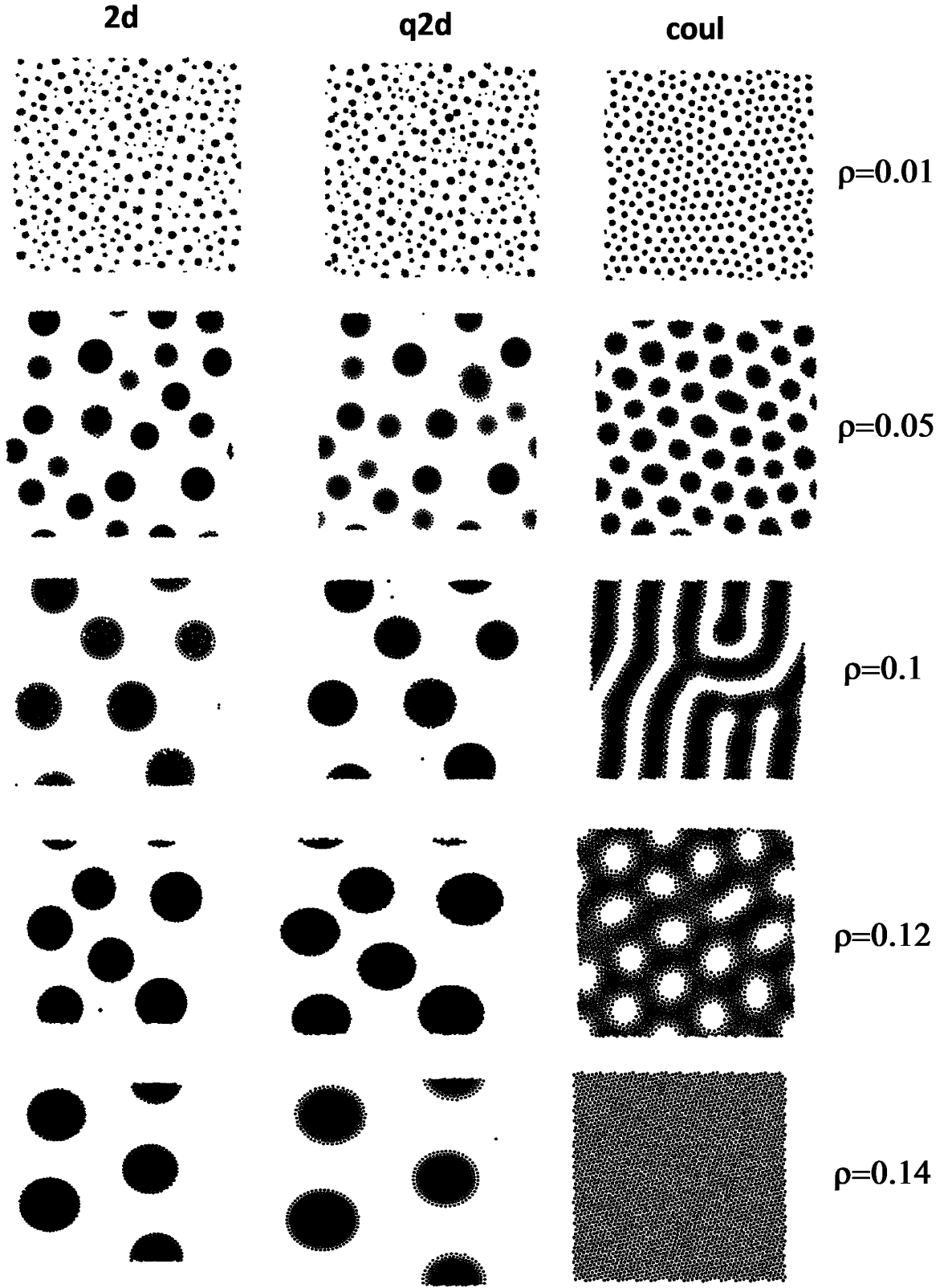}%

\caption{\label{snap-low-den} Snapshots of configuration of the
2d, q2d and coul systems at $T=0.001$ and densities up to
$\rho=0.14$. Further figures give more close views of these structures.}
\end{figure}

We started with investigation of the structure of three types of
systems: 2d, q2d and coul at $T=0.001$. Let us first consider the
system structure at low densities $\rho < 0.18$. Fig.
\ref{snap-low-den} shows snapshots of the systems at several
densities. Two major conclusions follow from these snapshots.
First, the structures of 2d and q2d are qualitatively identical.
Below we will address this question in more detail. This
conclusion confirms that transformation of a 2d system into q2d
does not influence the structure. The second conclusion indicates
that the structure of a coul system can be substantially different
from 2d and q2d. Combining it with the first one we conclude that
the changes of the structure are related to the long-range
interaction in the coul system.

Let us discuss in more detail the structures of the coul system at
different densities. At very low density $\rho=0.01$ the system
forms numerous small clusters. In Ref. \cite{bishop} it was
proposed to call this phase  a clump phase. Fig. 2 enlarges  part
of the snapshot from Fig. \ref{snap-low-den}. One can see that the
clusters can have different internal structures which also depend
on  cluster size. The small clusters mostly have triangular
symmetry. The larger one can have not so well defined crystal-like
ordering. Fig. 2 (b) shows the distribution of the size of the
clusters in the system. One can see that the smallest clusters
consist of 4 particles while the largest one  of 27. The most
probable cluster size is 10 or 12 vortices (two peaks of the
distribution functions). The vv-RDF demonstrates the first
extremely high peak at $r=4$, the second peak is located at
$r=7.46$ the third small peak at $r=10.57$ (Fig. 2 (c)). At a
distance of about $r=14$ a gap with an almost zero value of $g(r)$
is observed, so this distance can be considered as the average
size of the clusters. The next peak of the vv-RDF coincides with
the peak of the cc-RDF, i.e. here it already determines the
correlation between the particles of different clusters. The
clusters form a kind of liquid.

\begin{figure} \label{coul-rho001a}
\includegraphics[width=6cm,height=4cm]{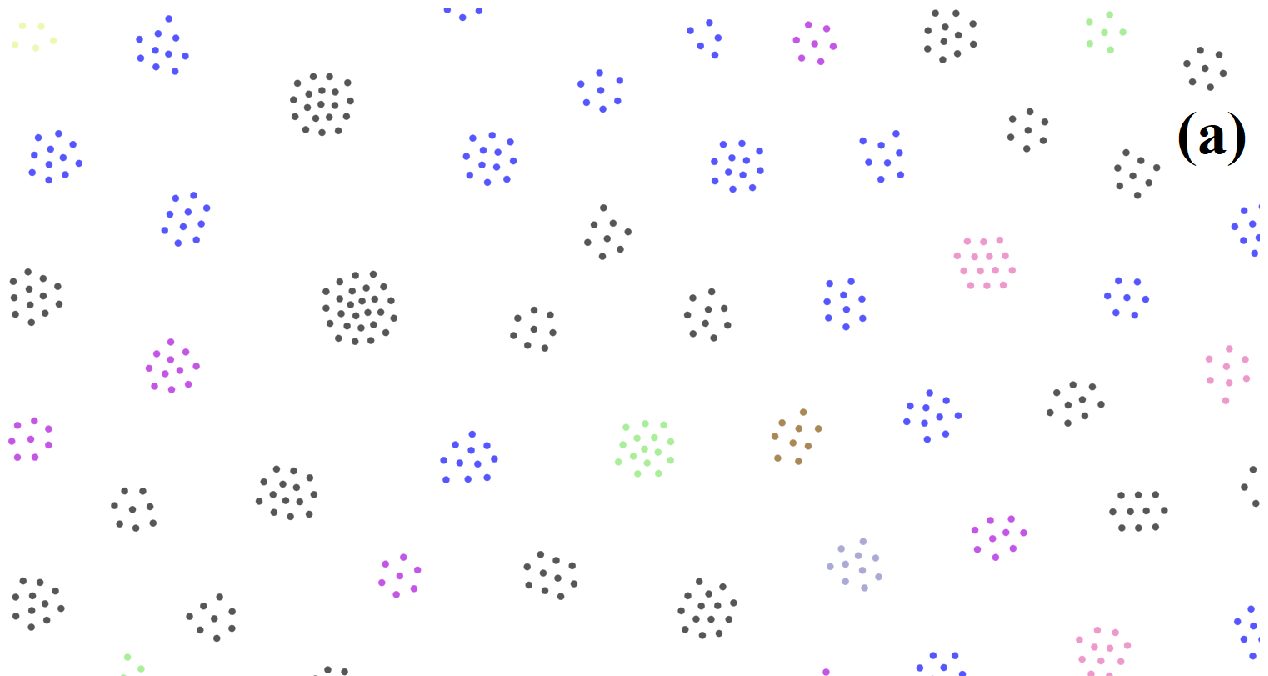}%

\includegraphics[width=6cm,height=6cm]{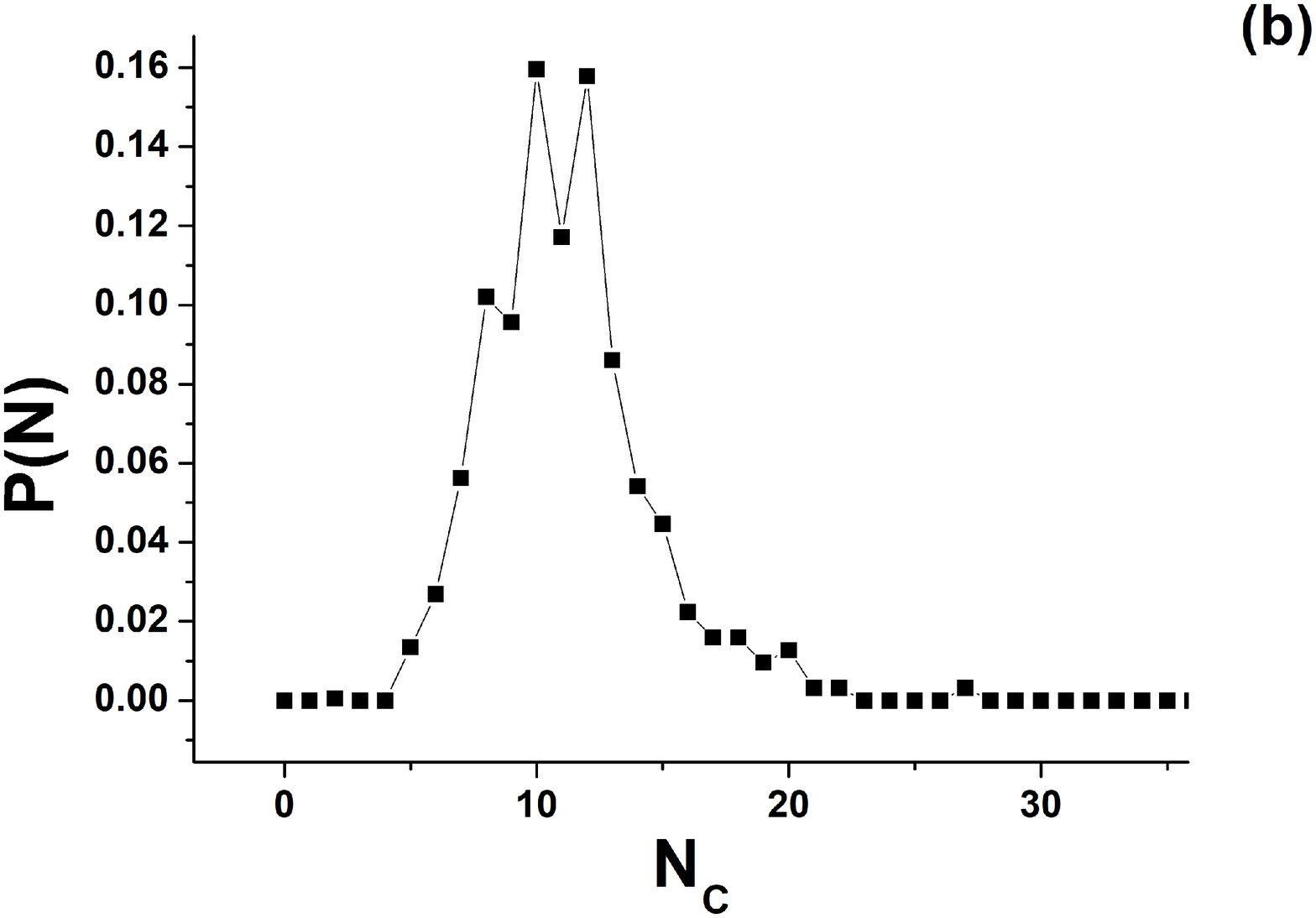}%

\includegraphics[width=6cm,height=6cm]{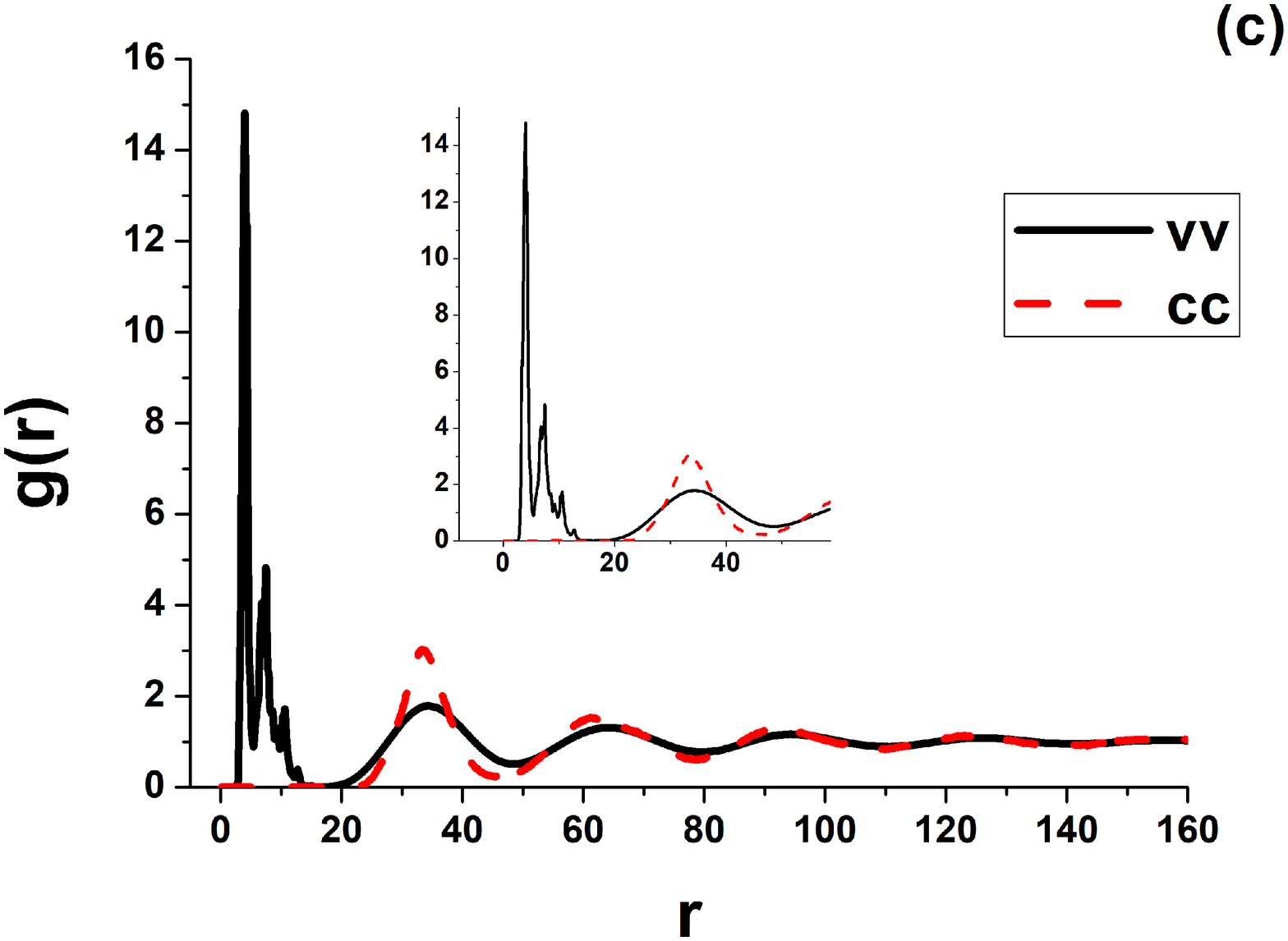}%

\caption{(a) An enlarged snapshot of a coul system at $T=0.001$
and $\rho=0.01$; (b) the distribution of the cluster size in the
same system; (c) the vv- and cc- RDFs of the same system. The
inset enlarges the part of the plot at low r.}
\end{figure}

If the density is increased to $\rho=0.05$ then the coul system
forms a cluster phase with larger clusters (Fig.
\ref{snap-low-den}). Fig. 3 (a) enlarges  part of the snapshot of
the system. Panel (b) shows cluster size distribution. The
smallest cluster consists of 50 vortices while the largest one of
98. The highest peak corresponds to 68 particles in a cluster.
From the RDFs of this system (Fig. 3 (c)) we can conclude that the
internal structure of the clusters does not show any kind of
crystalline order observed in some lower density clusters.
Importantly, the outer shell of a cluster is separated from the
other particles of the same cluster by slightly greater distance
compared with the internal particles. This shell is responsible
for the second vv-RDF peak  at $r=4.79$ (see the inset of Fig. 3
(c)). The first cc-RDF peak is located at $r=40$. Since the size
of the clusters is comparable to the distance between them the
vv-RDF does not go to zero at  distances exceeding the cluster
size. We can roughly estimate the size of the clusters as a
minimum in the vv-RDF which is located at $r=20.1$. Importantly,
at this density a triangular superstructure of clusters appears
which is visible from the snapshots and cc-RDF. However, larger
systems are required for detailed investigation of this
superstructure.

\begin{figure} \label{coul-rho005a}
\includegraphics[width=6cm,height=4cm]{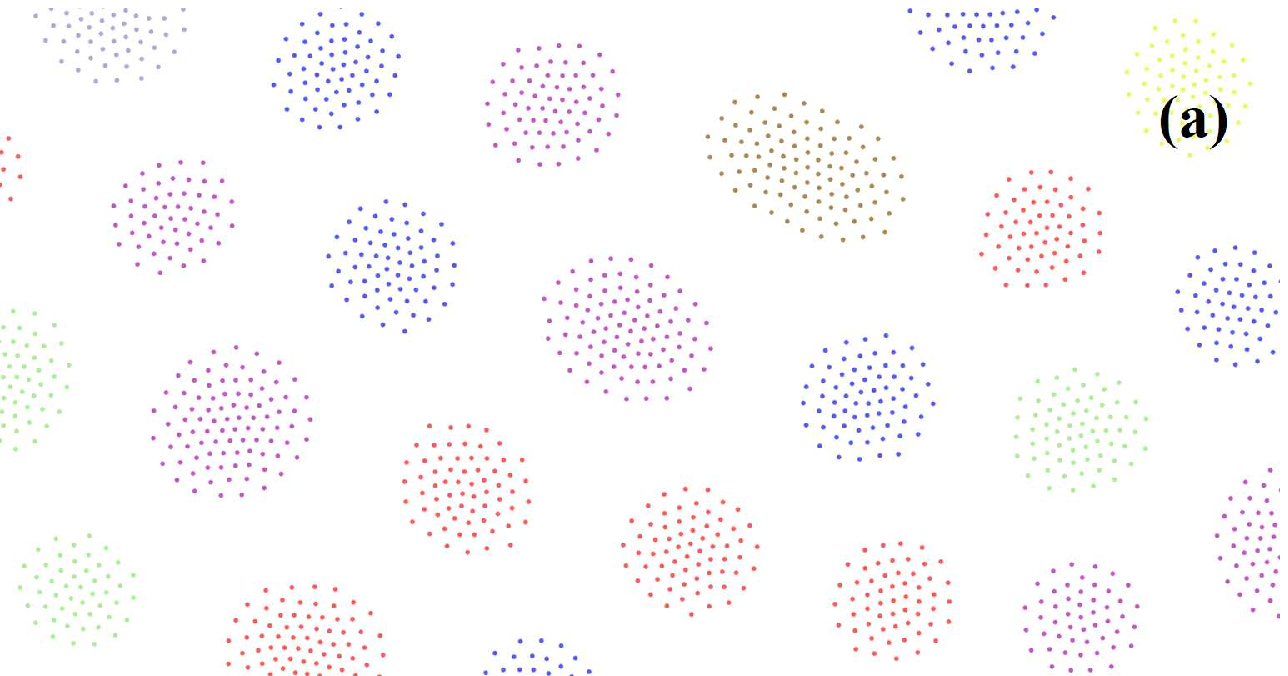}%

\includegraphics[width=6cm,height=6cm]{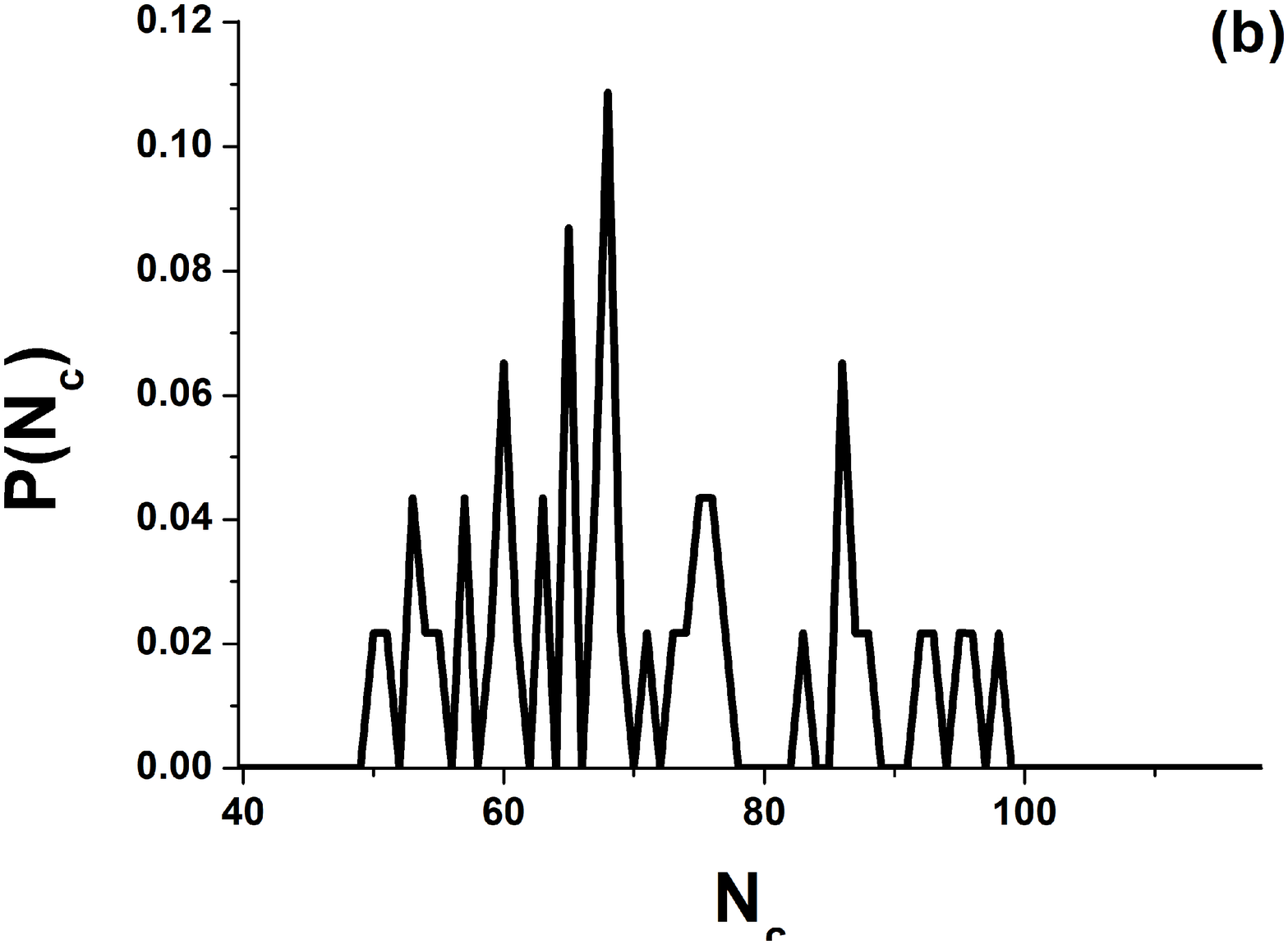}%

\includegraphics[width=6cm,height=6cm]{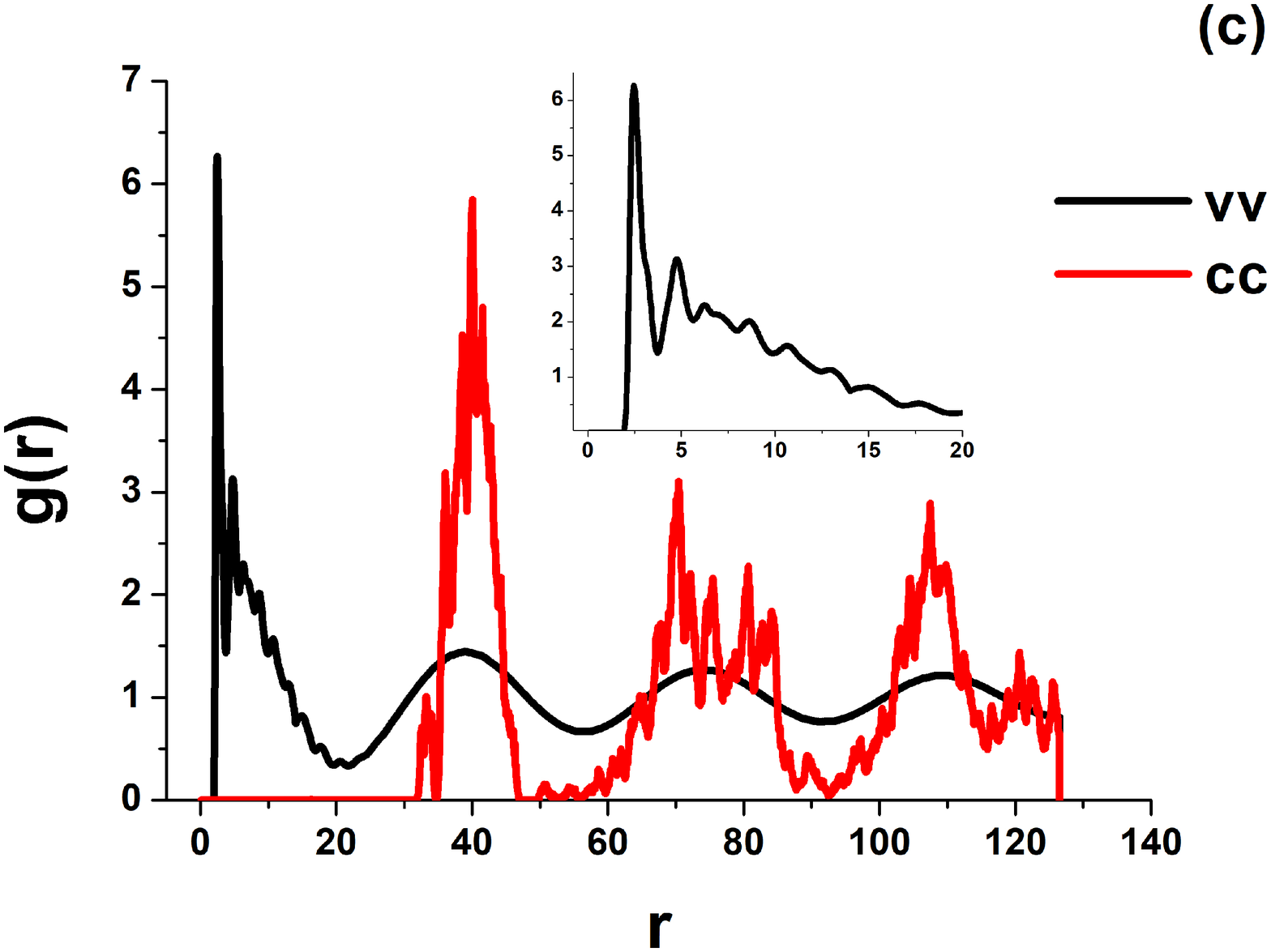}%

\caption{(a) An enlarged snapshot of the coul system at $T=0.001$
and $\rho=0.05$; (b) the distribution of the cluster size in the
same system; (c) the vv- and cc-RDFs of the same system. The inset
enlarges the small r part of the vv-RDF.}
\end{figure}

In the case of density $\rho=0.1$ the structure of the system
changes (Fig. \ref{snap-low-den}). At this density we observe a
single cluster percolating through the whole system. We propose to
call this phase a "Nazca phase" due to its similarity to the nazca
lines in South America. Fig. 4 (a) enlarges  part of the snapshot
of the nazca phase. One can see that the internal part of the
cluster has a triangular structure. At the same time, as in the
clusters at $\rho=0.05$ the outer shell of the vortices is located
at greater distance from the internal layers than the lattice
constant of the crystal. Fig. 4 (b) demonstrates the vv-RDF of
this system (There is no cc-RDF because the whole system is a
single cluster). One can see modulations on the RDF. The inset
shows the low r part of the RDF. The first peaks appear at
distances 2.4, 4.21, 4.8 and 6.43, which corresponds to a
triangular solid.

\begin{figure} \label{coul-rho01}
\includegraphics[width=6cm,height=4cm]{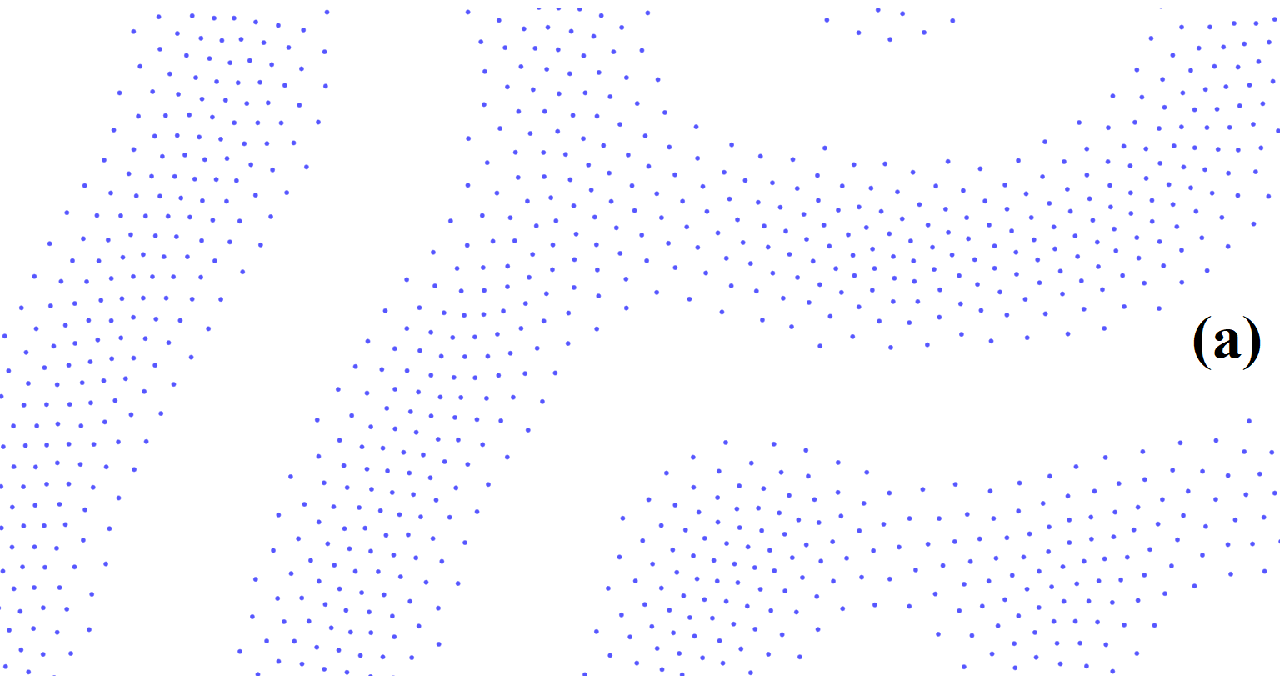}%

\includegraphics[width=6cm,height=6cm]{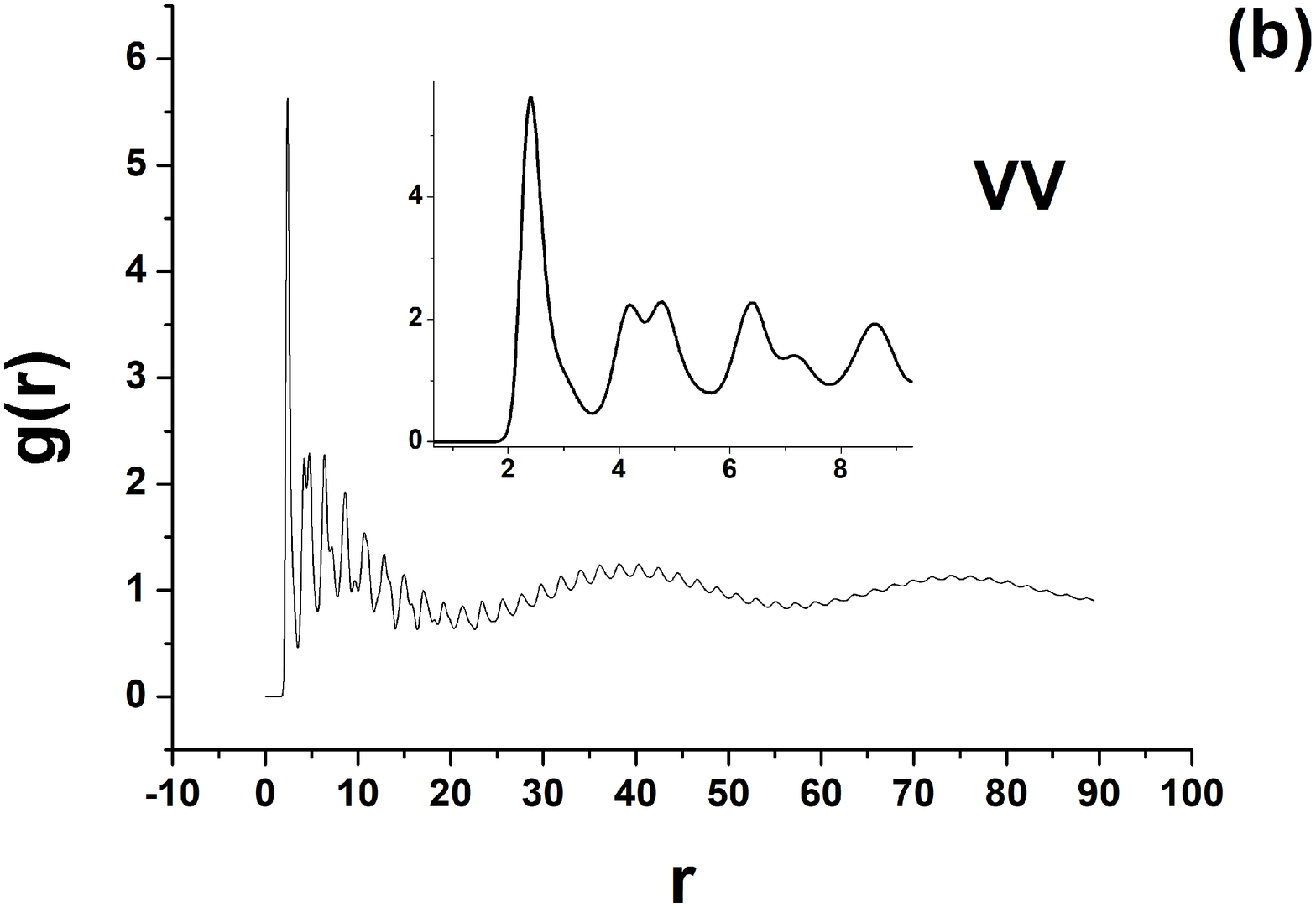}%

\caption{(a) An enlarged snapshot of the coul system at $T=0.001$
and $\rho=0.1$; (b) the vv- RDFs of the same system. The inset
enlarges the low r part of the RDF.}
\end{figure}

A further increase in density leads to inverting of the structure.
While at low densities we observe clustering of the system, at
density $\rho=0.12$ the coul system demonstrates a structure with
holes which in Ref. \cite{bishop} was called anti-clump. We
propose to call it a "cheese phase". A similar structure was
observed in Ref. \cite{meng3} for the same system, but with
slightly different parameters of the potential and without
consideration of the long range interaction. The vortices of the
cheese phase are ordered in the triangular phase which can be seen
from the RDF shown in Fig. 5. At the same time it seems that the
holes also form a triangular superstructure. However, this
question requires further investigation with larger systems.

\begin{figure} \label{coul-rho012}
\includegraphics[width=6cm,height=6cm]{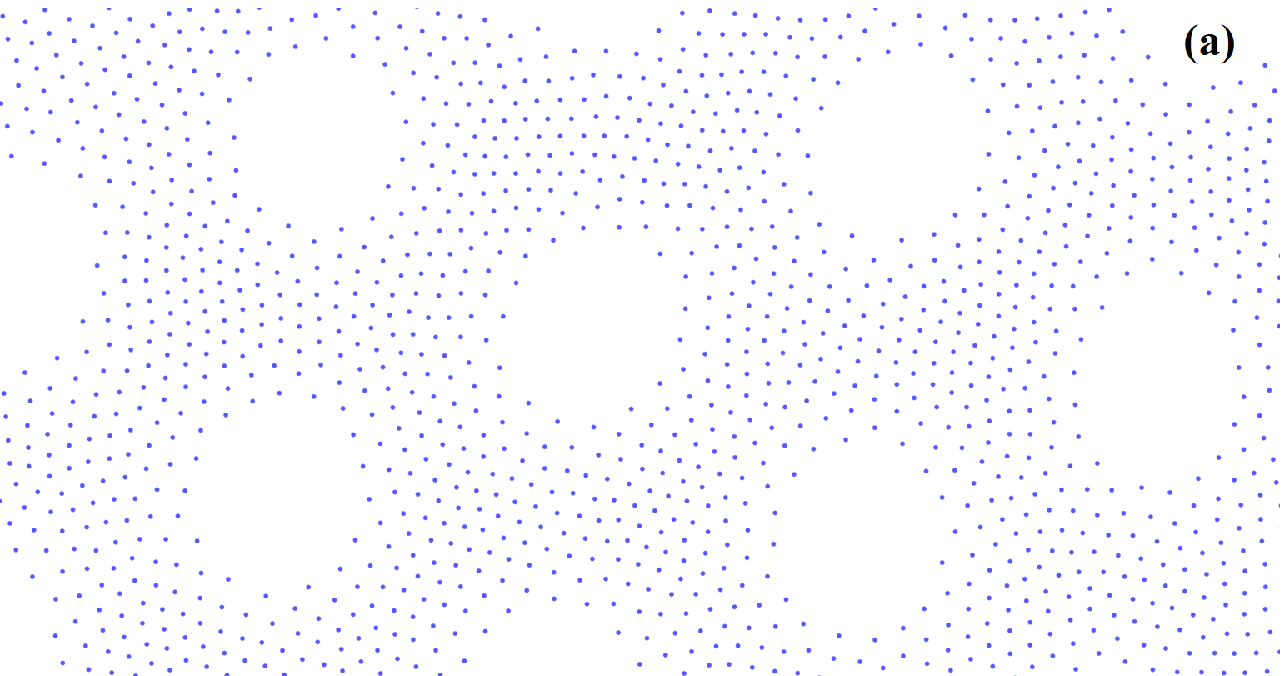}%

\includegraphics[width=6cm,height=6cm]{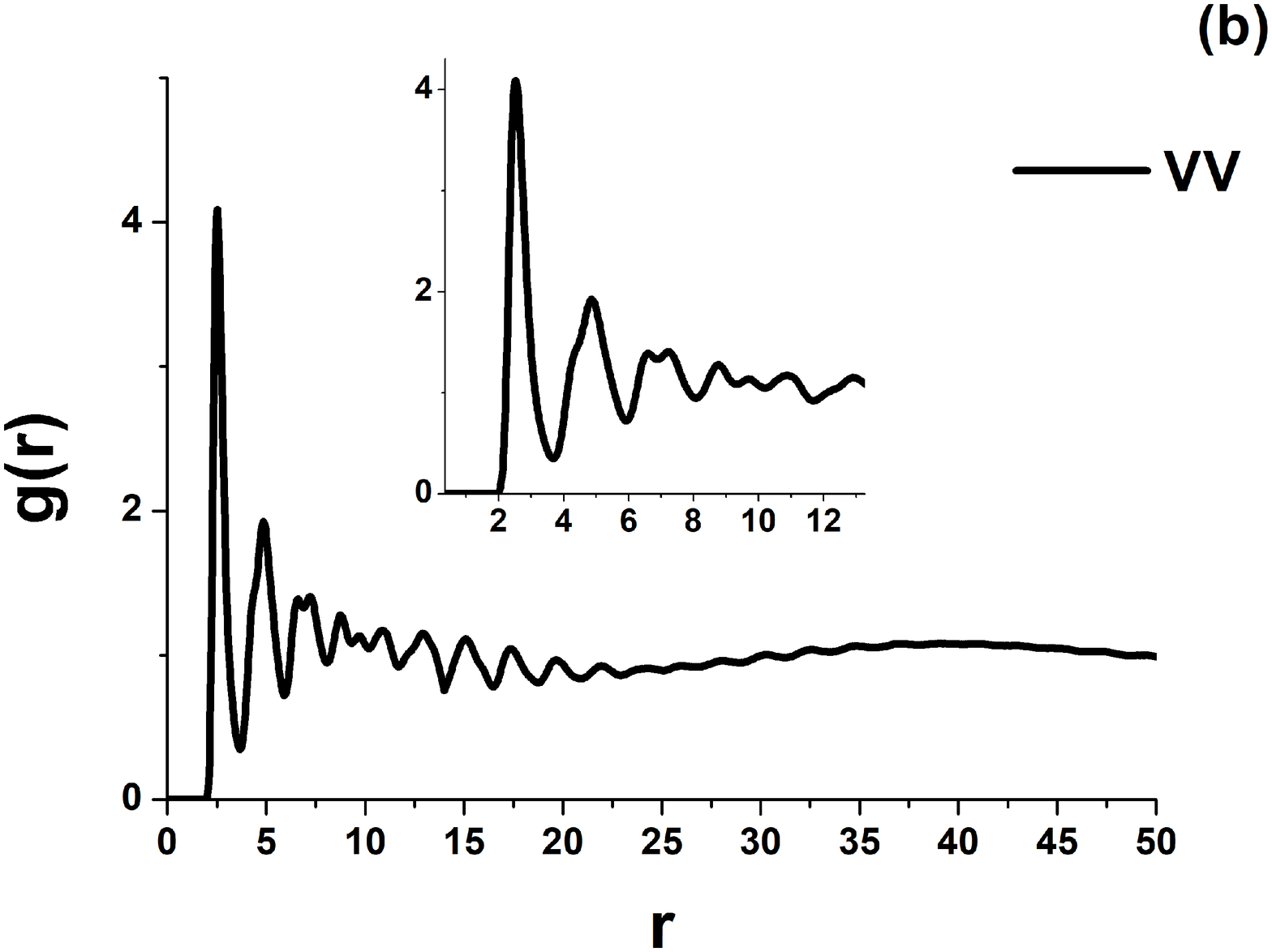}%

\caption{The vv-RDFs of the coul system at $\rho=0.12$ and
$T=0.001$. The inset enlarges the low r part of the plot.}
\end{figure}

We finish the discussion of the coul system at low densities by
investigation of the system at $\rho=0.14$. From the snapshot in
Fig. \ref{snap-low-den} one can see that this system forms a usual
triangular lattice. This conclusion is supported by the vv-RDFs of
the system given in Fig. 6.

At  densities $\rho=0.16$ and $\rho=0.18$ the coul system
preserves a triangular lattice. Because of this we do not show the
results for these densities.


\begin{figure} \label{coul-rho014}
\includegraphics[width=6cm,height=6cm]{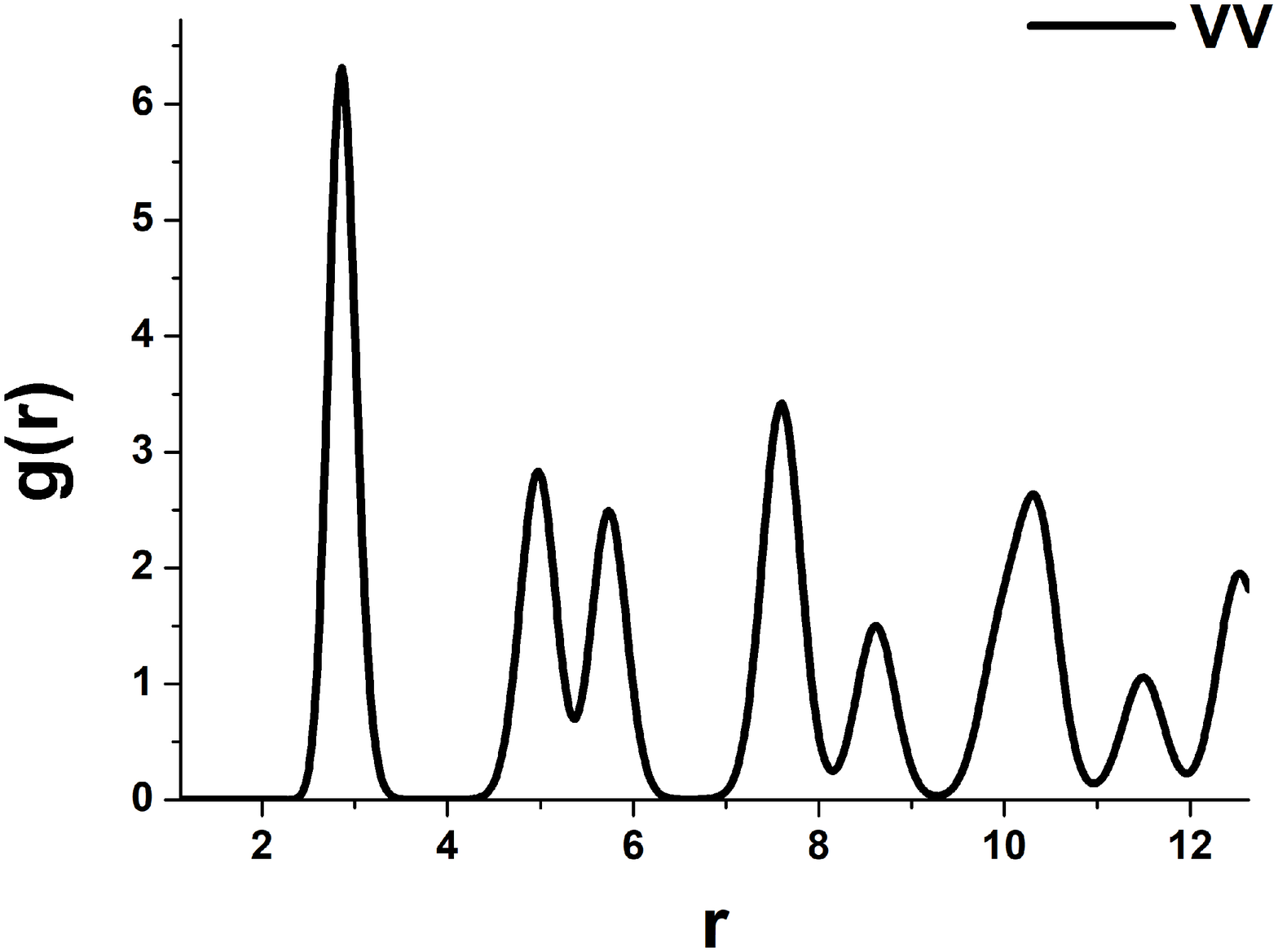}%

\caption{The vv-RDFs of the coul system at $\rho=0.14$ and
$T=0.001$.}
\end{figure}

Now we turn to the description of  2d and q2d systems at the same
densities. As mentioned above, the structure of these systems
appears to be identical which can be seen from Fig.
\ref{snap-low-den}. At the same time the behavior of these systems
is more simple compared with the case of the coul one. 2d and q2d
systems form clusters at all densities $\rho<0.18$. As the density
increases the size of the clusters also increases. At density
$\rho=0.01$ the clusters are small and some of them demonstrate a
triangular internal structure. At higher densities the clusters
consist of a triangular core and several layers above it. We will
consider the internal structure of these clusters below.
Importantly, the 2d and q2d systems demonstrate unusual dynamics
of vortices. The systems rapidly split into  fast and slow parts.
Some of the vortices move extremely actively, while others quickly
condense into  clusters and move slowly. As soon as the clusters
grow the fraction of  slow particles will increase. However, some
of the particles do not fall into  clusters for very long time.
Because of this we observe many single vortices in the cluster
phases of 2d and q2d systems. We expect that in the limit of
infinite time all vortices will belong to clusters. It is also
worth noting that such a cluster phase was observed in Refs.
\cite{meng1,meng2,meng3}, but the authors of these publications
considered only the ground state of the system and therefore they
did not observe such peculiar dynamic behavior of  vortices.


\subsection{Intermediate and high densities}

\begin{figure}
\includegraphics[width=18cm,height=22cm]{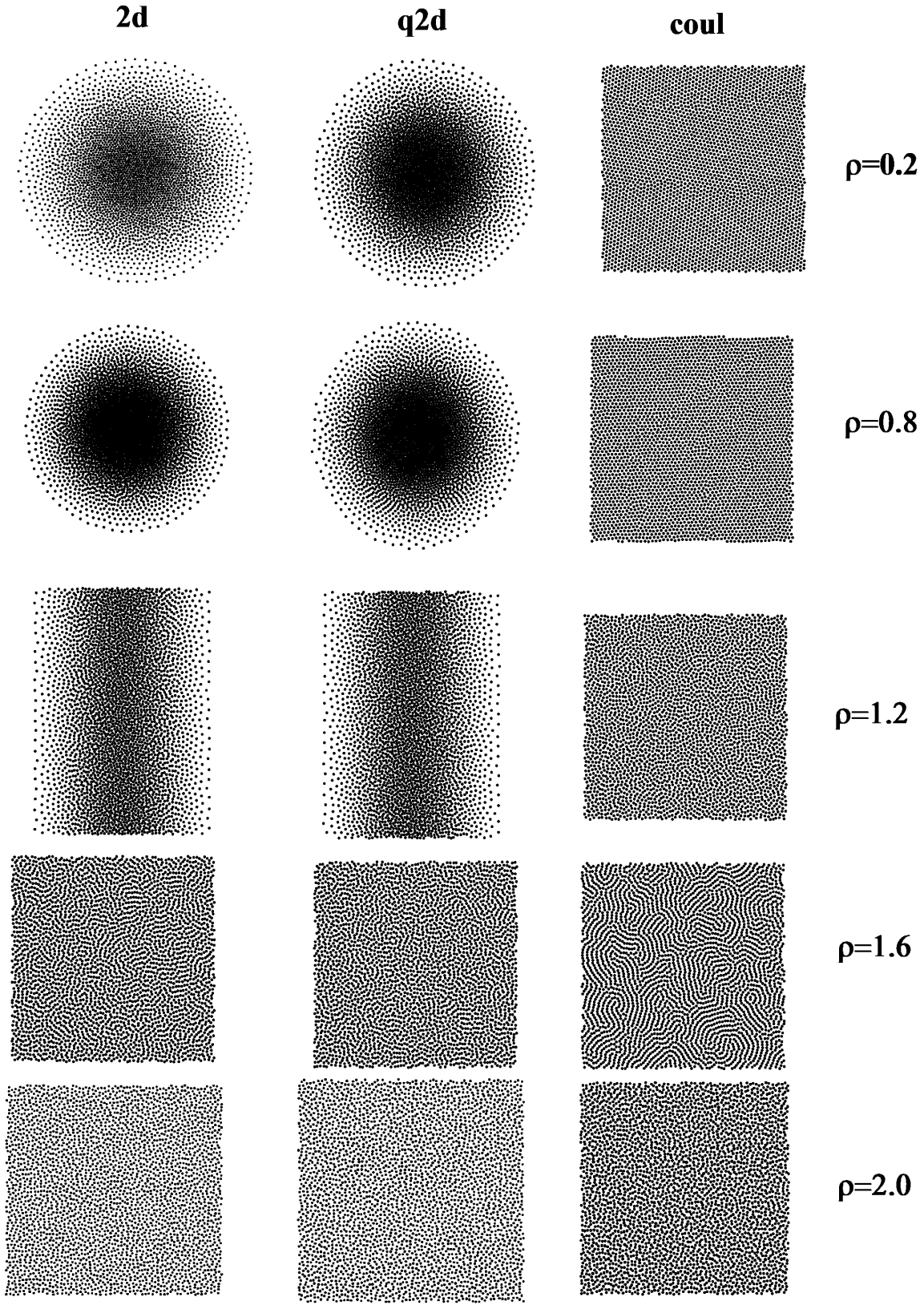}%

\caption{\label{snap-high-den} Snapshots of the configuration of
2d, q2d and coul systems at $T=0.001$ and densities up to
$\rho=2.0$. The different colors mark different clusters in the
systems. In some cases the number of clusters exceeds the number
of colors used and different clusters can have the same color.
Periodic boundary conditions are used, therefore the clusters can
partially be located at  different edges of the image.}
\end{figure}


Fig. \ref{snap-high-den} shows snapshots of  2d, q2d and coul
systems at five different densities. Here we show only the
densities at which some changes in the structure take place.


We start the discussion again from the coul system. This system
preserves a triangular crystal phase up to  density $\rho=1.2$
where it becomes disordered. At density $\rho=1.6$ the system
forms a stripe liquid which was earlier observed in several
core-softened systems \cite{camp1,camp2,malescio,nk} and in the
experiment with colloidal films in a magnetic field
\cite{dobnikar}. Finally, at density $\rho=2.0$ the stripe phase
collapses into another disordered phase. Apparently, the high
density limit of this phase should be a triangular lattice,
therefore higher densities are out of the scope of the present
work.

Let us now turn to 2d and q2d systems. The densities studied in
this part of the paper correspond to the range of densities
investigated in Ref. \cite{meng3}, therefore we can make a
comparison of the results. However, one should remember that in
Ref. \cite{meng3} only the ground state of the system was studied,
while we perform simulations at finite temperatures.

From Fig. \ref{snap-high-den} one can see that the structures of
2d and q2d systems are again very similar. In agreement with Ref.
\cite{meng3} the systems form a circular cluster at  density $0.2$
which transforms to a ribbon at $\rho=1.2$. It is important that
the cluster at $\rho=1.2$ does not occupy the whole simulation
box. The whole system can be described as a ribbon surrounded by a
vacuum. However, at $\rho=1.6$ we observe a stripe phase. This is
in contradiction to Ref. \cite{meng3} where a cluster phase is
reported for densities up to $\rho=2.5$. This contradiction can be
related to the difference in temperature: while in Ref.
\cite{meng3} only ground states were considered we perform
investigations at finite temperature.


We again see that the contribution of long-range interactions
completely changes the structure of the system. The origin of this
effect can be explained by exploration of the RDFs of  2d, q2d and
coul systems $g(r)$. Fig. 8 shows $g(r)$ of all three types of
systems at $\rho=0.2$ and $T=0.001$.


In the case of $\rho=0.2$ the difference in the radial
distribution functions of 2d (or q2d) and coul systems is
dramatic. Importantly, the first peak appears at $r_1=0.55$ in the
2d system, $r_1=0.58$ in the q2d and $r_1=2.4$ in the coul one. It
means that a long range interaction changes the characteristic
distances in the system. Indeed the nature of the interaction is
different in the case of short and long-range ones. In the former
case only several particles which are within the cut-off distance
to the one under consideration interact with it. In the latter
case each particle interacts with the whole system, which can
substantially shift the energy and the characteristic distances in
the system.

\begin{figure} \label{rdf-r02-all}
\includegraphics[width=6cm,height=6cm]{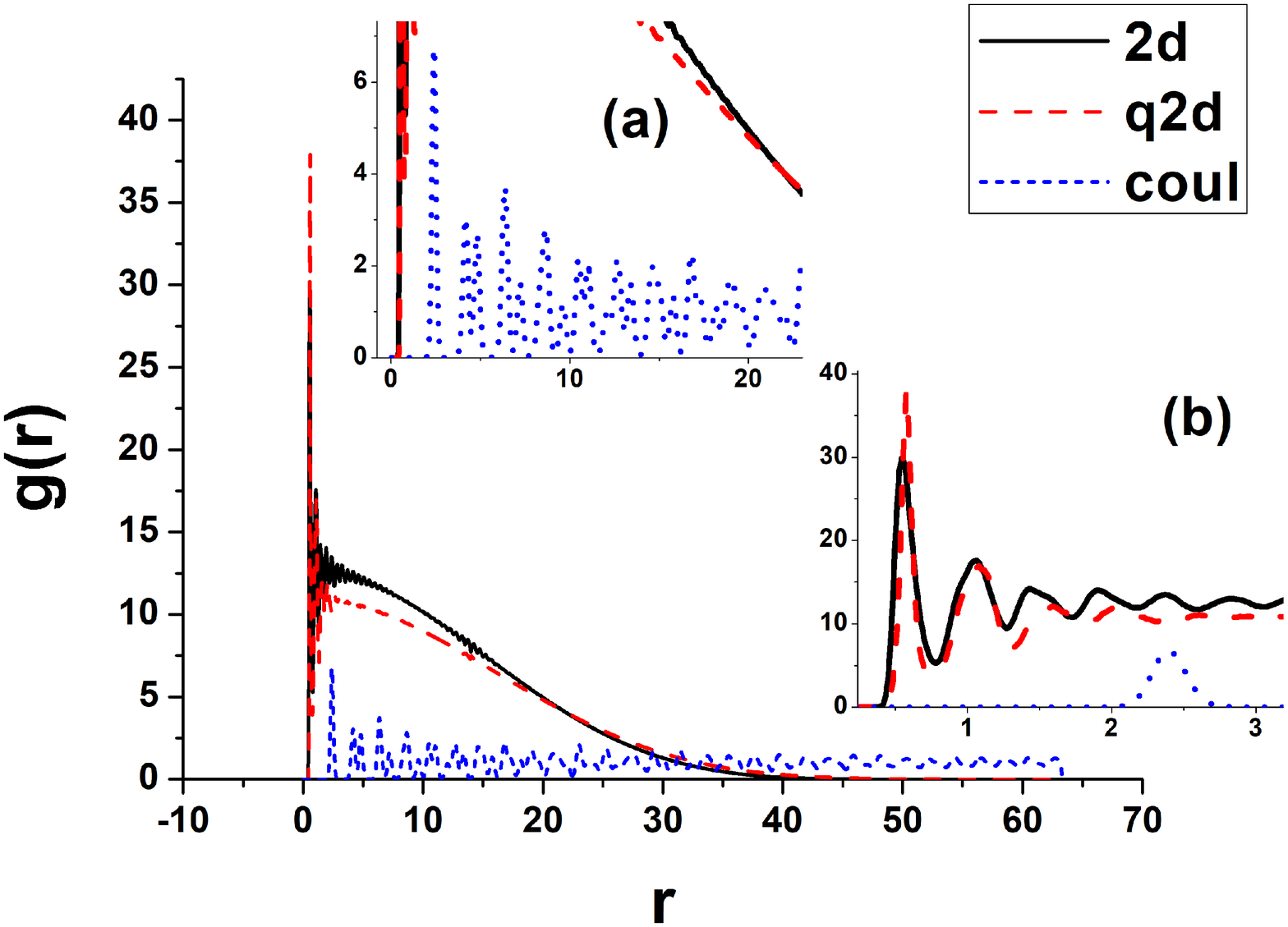}%

\caption{The radial distribution functions of 2d, q2d and coul
systems at $\rho=0.2$ and $T=0.001$. Inset (a) enlarges the RDF of
the coul system. Inset (b) enlarges the low r region.}
\end{figure}

The difference in $g(r)$ becomes less pronounced as density
increases. The first g(r) peaks  of 2d, q2d and coul systems
become closer to each other, and for densities above $1.8$ the
RDFs of all three types of  systems become indistinguishable which
is illustrated in Fig. 9. It means that while a long range
interaction determines the structural properties at low densities
it becomes much less significant at high densities.

\begin{figure} \label{rdf-02-12-18}
\includegraphics[width=6cm,height=6cm]{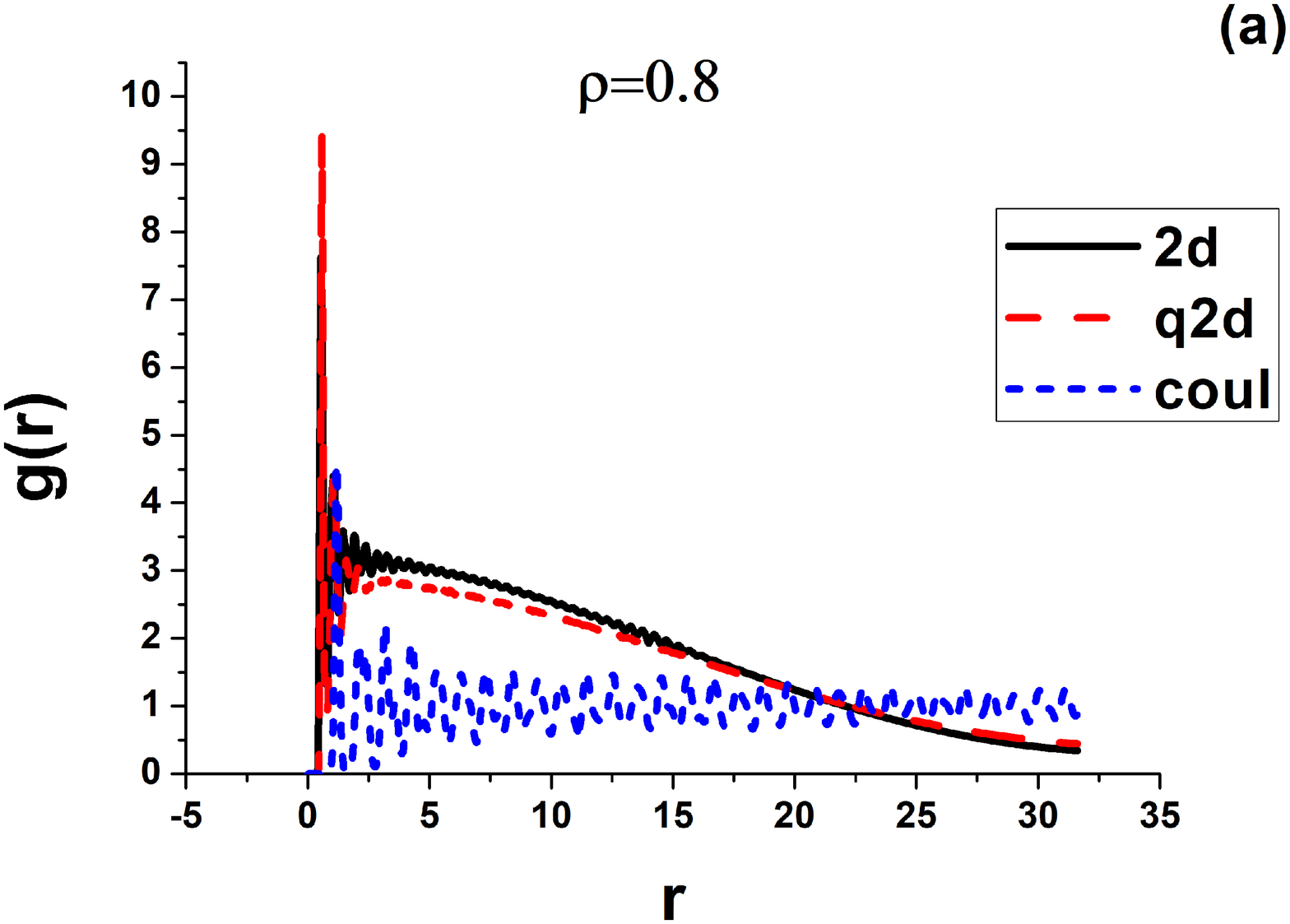}%

\includegraphics[width=6cm,height=6cm]{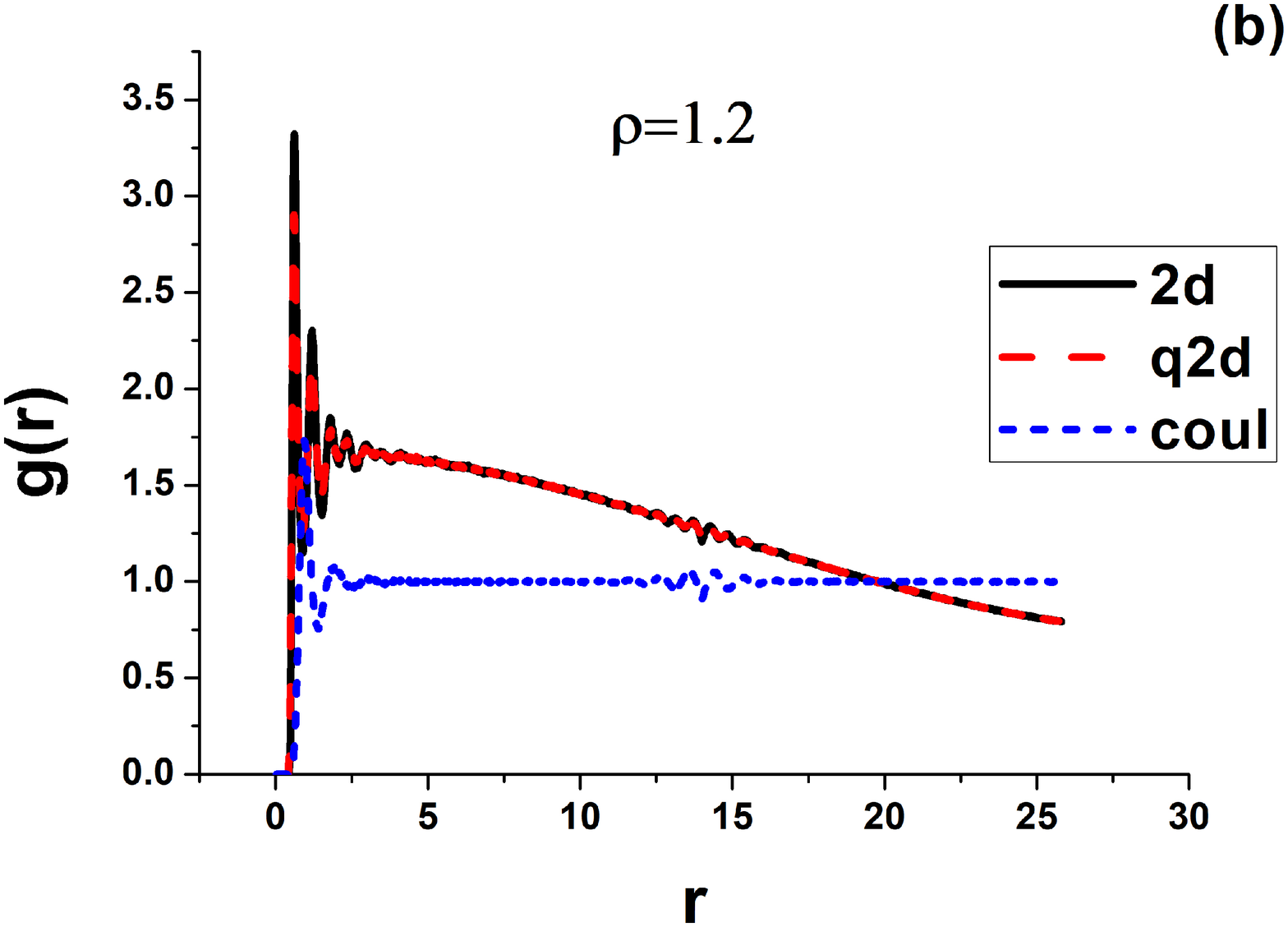}%

\includegraphics[width=6cm,height=6cm]{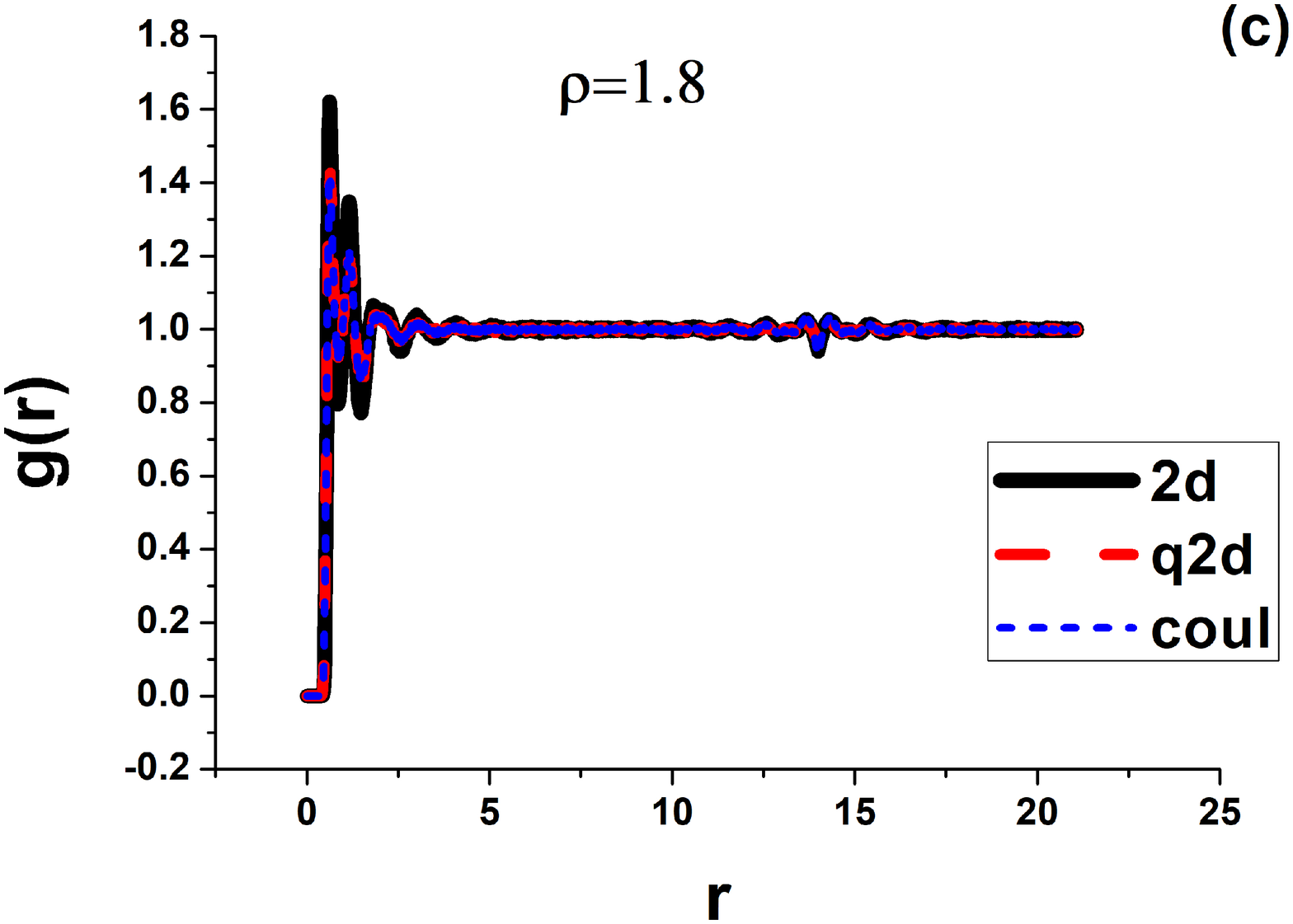}%

\caption{The radial distribution functions of 2d, q2d and coul
systems at $T=0.001$ and (a) $\rho=0.8$; (b) $\rho=1.2$ (c)
$\rho=1.8$.}
\end{figure}


\begin{figure}\label{rho02-clust}
\includegraphics[width=8cm]{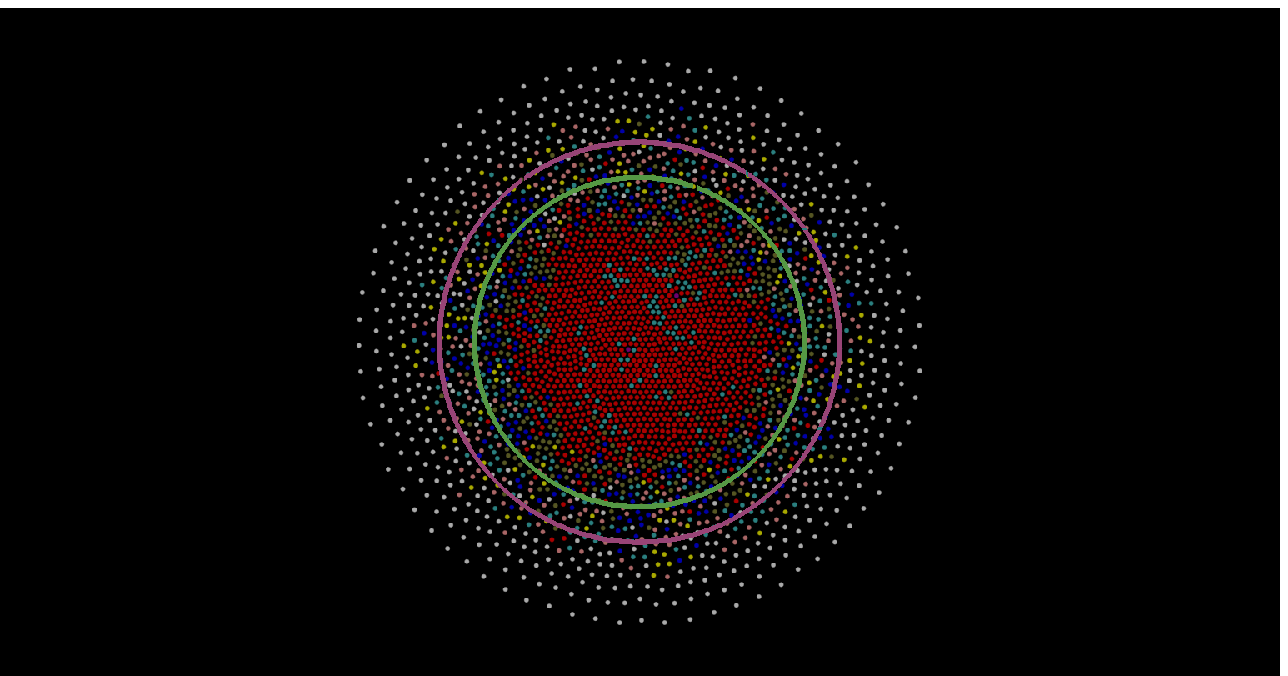}%

\caption{ A snapshot of a cluster at $\rho=0.2$ and $T=0.001$. The
particles are colored according to their coordination number. The
inner circle denotes the boundary of the inner crystalline core of
the cluster. The outer circle bounds the region where the distance
to  neighboring particles exceeds the first RDF minimum.}
\end{figure}

Let us consider the cluster phase of 2d systems in more detail.
Fig. 10 shows a snapshot of the 2d system with $\rho=0.2$. The
particles are colored according to the number of neighbors they
have. The number of neighbors was defined as the number of
particles which were closer to a given one than the first minimum
of g(r). One can see that the number of neighbors can have values
from zero to 6. From the snapshot one can see that close to the
center of the cluster the particles are arranged in a triangular
lattice. However, going from the cluster center to the edge the
distance between  particles becomes greater. The border particles
of the cluster form a circle with radius $R=23.718$. In order to
characterize the internal structure of the cluster we calculate
the radial distributions of  density, order parameter $\psi_6$ and
number of nearest neighbors.

Parameter $\psi_{6,i}$ where i is the particle number is defined
as

\begin{equation}
  \psi_{6,i}=\sum_{j=1}^{NN_i} exp(6i\theta_{ij}),
\end{equation}
where the j is one of the nearest neighbors of i and $\theta_{ij}$
is the angle between  vector ${\bf r_{ij}}$ and an arbitrary
direction. The global order parameter in some region is defined as

\begin{equation}
\psi_6=\sum_{i} \psi_{6,i},
\end{equation}
where $i$ denotes  particles belonging to the region of interest.

The radial distributions of different quantities are defined as
follows. From analysis of the cluster outer shell we find that it
has a circular shape. If we place the origin of the coordinate
frame in the center of this circle we can determine the circular
layer bounded by concentric circles with radius r and $r+dr$. The
area of such a layer is $S(r+dr,r)=\pi (r+dr)^2 - \pi r^2$. We
define the radial density of a cluster as the number of particles
inside such a layer divided by  area $S(r)$. The sum of
$\psi_{6,i}$ for all particles in the layer divided by this number
of particles is the radial distribution of $\psi_6$. The radial
distribution of the nearest neighbors is calculated in the same
way.

The radial density has a very sharp peak at the origin. Two
effects are responsible for this phenomenon. First, there are many
particles next to the center of the circle and, second, the area
of the layer from zero to $dr$ is extremely small. The
distribution of order parameter $\psi_6$ shows a plateau at
radiuses from zero up to 8. After that it decays monotonically and
becomes negligible at $r=16.3$. We define the crystalline segment
of the cluster as a part with $\psi_6 \geq 0.4$. In this case the
radius of the crystalline part is $r_{cr}=13.93$. The outer layers
where the particles have no nearest neighbors extends from
$r=16.3$ up to $23.718$, i.e. the width of this outer shell is
$\Delta =7.418$.

\begin{figure}
\includegraphics[width=8cm]{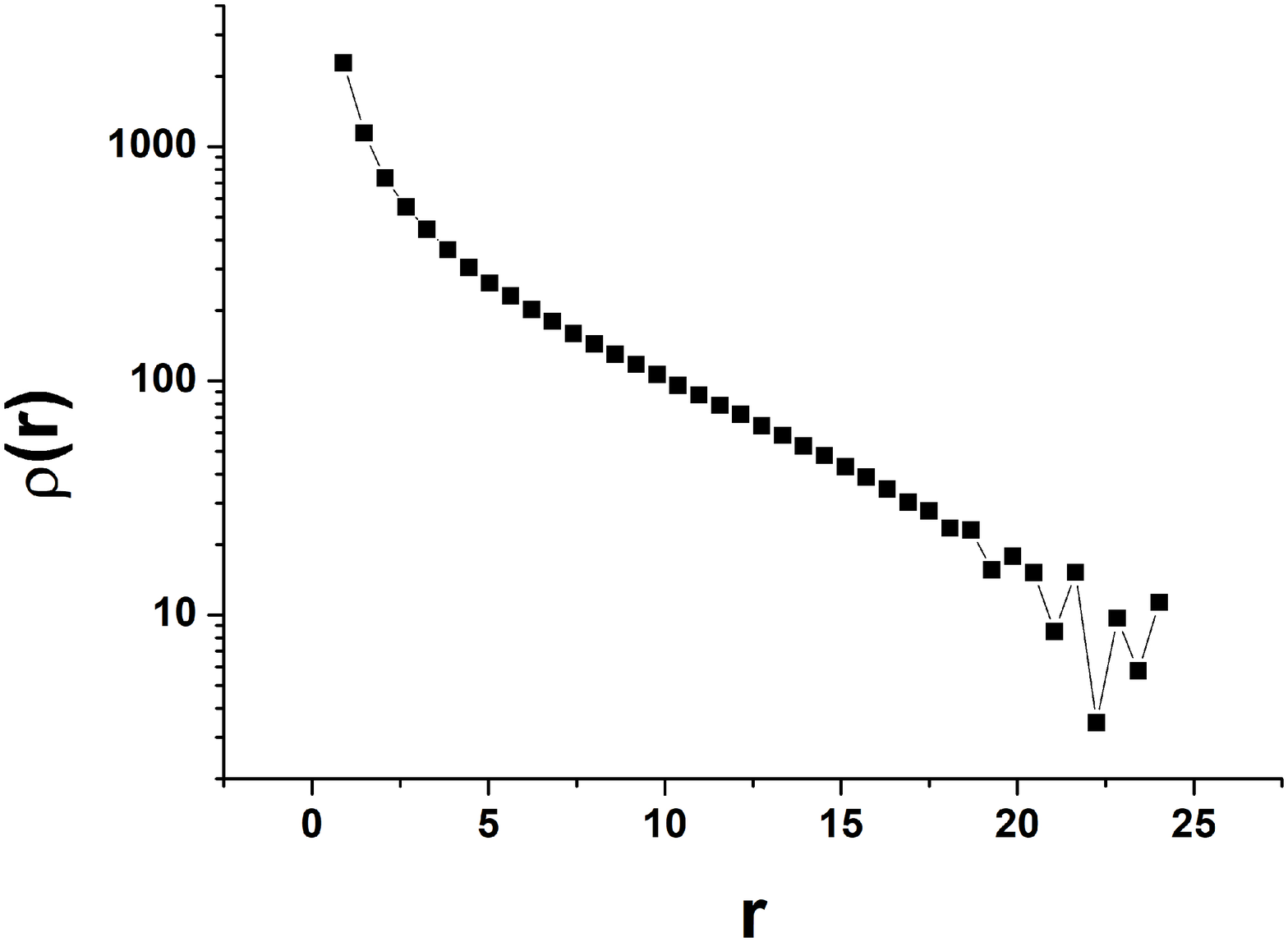}%

\includegraphics[width=8cm]{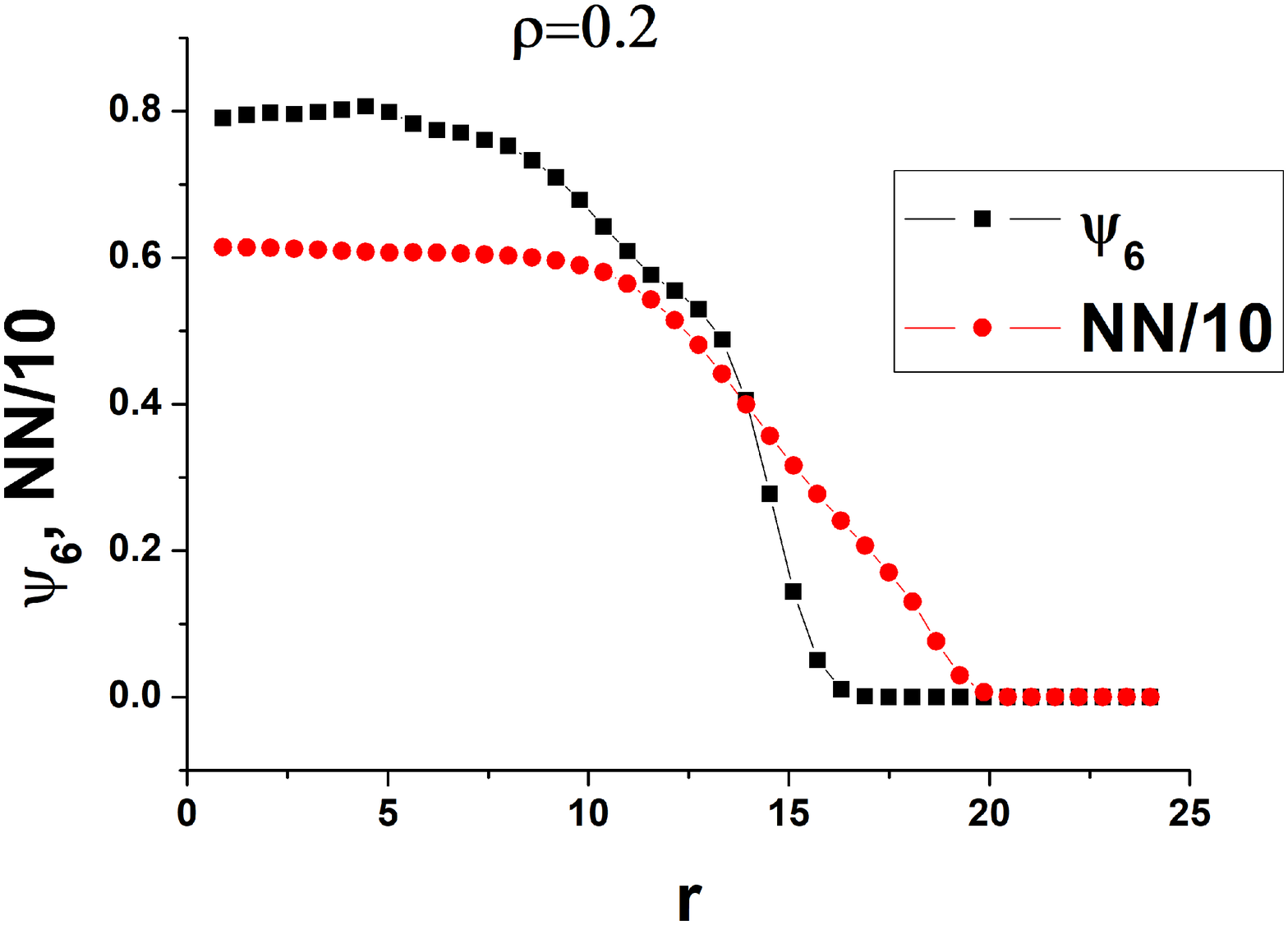}%

\caption{\label{fig:rho02-rpn} (a) The radial density distribution
of the 2d phase cluster at density $\rho = 0.2$ and $T=0.001$. (b)
The radial distribution of order parameter $\psi_6$ and the number
of nearest neighbors NN at the same point. NN is divided by 10 in
order to place both curves on the same plot.}
\end{figure}

The same analysis is performed for the other clusters of the 2d
system for densities $0.4$, $0.6$, $0.8$ and $1.0$ (not shown in
the text).  Interestingly, the cluster size appears to be
independent of the density: the radial densities for all densities
of the system are identical except $\rho =1.0$ where the radial
density is a bit higher. Also the sizes of the cluster crystalline
core and outer zero neighbor shell appear to be identical for all
the cases except $\rho=1.0$. To be more specific, the radius of
the cluster is from $R=23.704$ to $R=23.74$, the inner crystalline
core is $r_{cr}=13.93$ and the outer shell $r_{shell}=7.4$ for the
densities from $0.2$ up to $0.8$ and $R=25.775$, $r_{cr}=13.9$,
$7.415$ for $\rho=1.0$.

For  densities $\rho=1.2$ and $1.4$ the system forms a ribbon-like
cluster instead of a circular one. However, the cluster internal
structure looks very similar to the circular case. The clusters
consist of an internal crystalline ribbon, a transition region and
an outer shell with a zero number of the nearest neighbors. For
the case of $\rho=1.2$ the crystalline core extends from $x=14.52$
to $x=27$ and the outer shells - from $x=0.0$ to $x=7$ and $x=34$
- $x=40.0$. In the case of $\rho=1.4$ the crystalline core is from
$x=11.17$ to $27.0$ and the outer shells are from $x=0.0$ to
$x=4.18$ and $x=34$ to $x=38.8$.

One can see that at low densities cluster size increases with
density while at  intermediate ones the whole system forms a
single cluster. However, it can be a finite size effect. Also, in
Ref. \cite{meng3} it was noticed that while small systems formed a
single cluster, larger ones could split into several clusters. In
order to check the influence of  system size we simulate much
larger systems of 50000 particles.

\begin{figure}
\includegraphics[width=8cm]{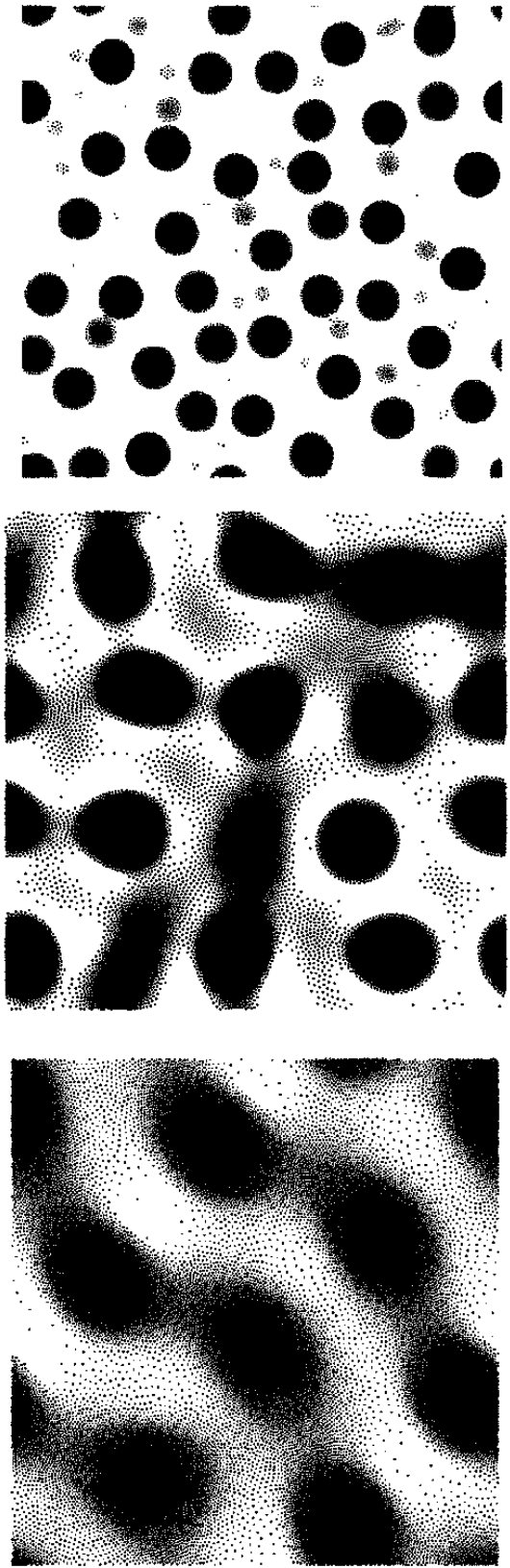}%

\caption{\label{2d-large} Snapshot of large 2d systems at
$\rho=0.2$ (top), $\rho=0.6$ (middle) and $\rho=1.2$ (bottom).}
\end{figure}

Fig. \ref{2d-large} shows snapshots of configurations of large 2d
systems at three densities: $\rho=0.2$, $0.6$ and $1.2$. At
$\rho=0.2$ the system apparently splits into numerous clusters.
However, the dynamics of the system is very slow. Even at
$\rho=0.2$ the system has not come into an equilibrium state yet.
One can see that small clusters are still included into  larger
ones. The dynamics of the systems at larger densities is even
slower. During the period of $1 \cdot 10^7$ steps we do not
observe any substantial changes in the structure. However, the
snapshots of both $\rho=0.6$ and $\rho=1.2$ resemble the initial
period of the system splitting into numerous spherical clusters.
From this observation we can conclude that the equilibrium state
at  densities $\rho=1.2$ and $1.4$ should rather be a state of
circular clusters, than a ribbon. We can suppose that the number
of vortices in circular clusters at these densities exceeds the
number of vortices in a small system (3200) which leads to the
formation of a ribbon-like cluster. For instance, at  density
$\rho=1.2$ the system has already split into 8 clusters which
means 6250 vortices in a cluster on average.

One should be careful to ensure that the system has reached a
state of equilibrium. For instance, the stripe and labyrinth
phases of Ref. \cite{njp} (see Fig. 7 of this reference) look
qualitatively similar to the phases of Fig. \ref{2d-large}. Due to
this we suppose that these phases are really not completely
equilibrated and finally they will change to the cluster phase.

\subsection{Cluster phase thermal stability }

Finally we estimated the thermal stability of the cluster phase in
three system types. In order to do this we increased the
temperature at two different densities: $\rho=0.01$ and
$\rho=0.2$. Small systems of 3200 vortices were used in this part
of the work. Fig. 13 shows the RDFs of 2d, q2d and coul systems at
$\rho=0.01$ and different temperatures. One can see that at low
temperature all systems demonstrate an extremely high first peak,
which corresponds to the distance between particles in the
clusters. When the temperature increases the peak becomes smaller
and finally the RDFs become almost flat which can be considered as
complete disappearance of ordering in the system. Moreover, at
some temperature the peak height drops dramatically. This
temperature can be considered as the "melting" temperature of
clusters. This "melting" temperature can be estimated as
$T_{2d}=0.008$ for the 2d system, $T_{q2d}=0.06$ for the q2d and
$T_{coul}=0.04$ for the coul one.

\begin{figure}\label{rdf-r001-t}
\includegraphics[width=8cm]{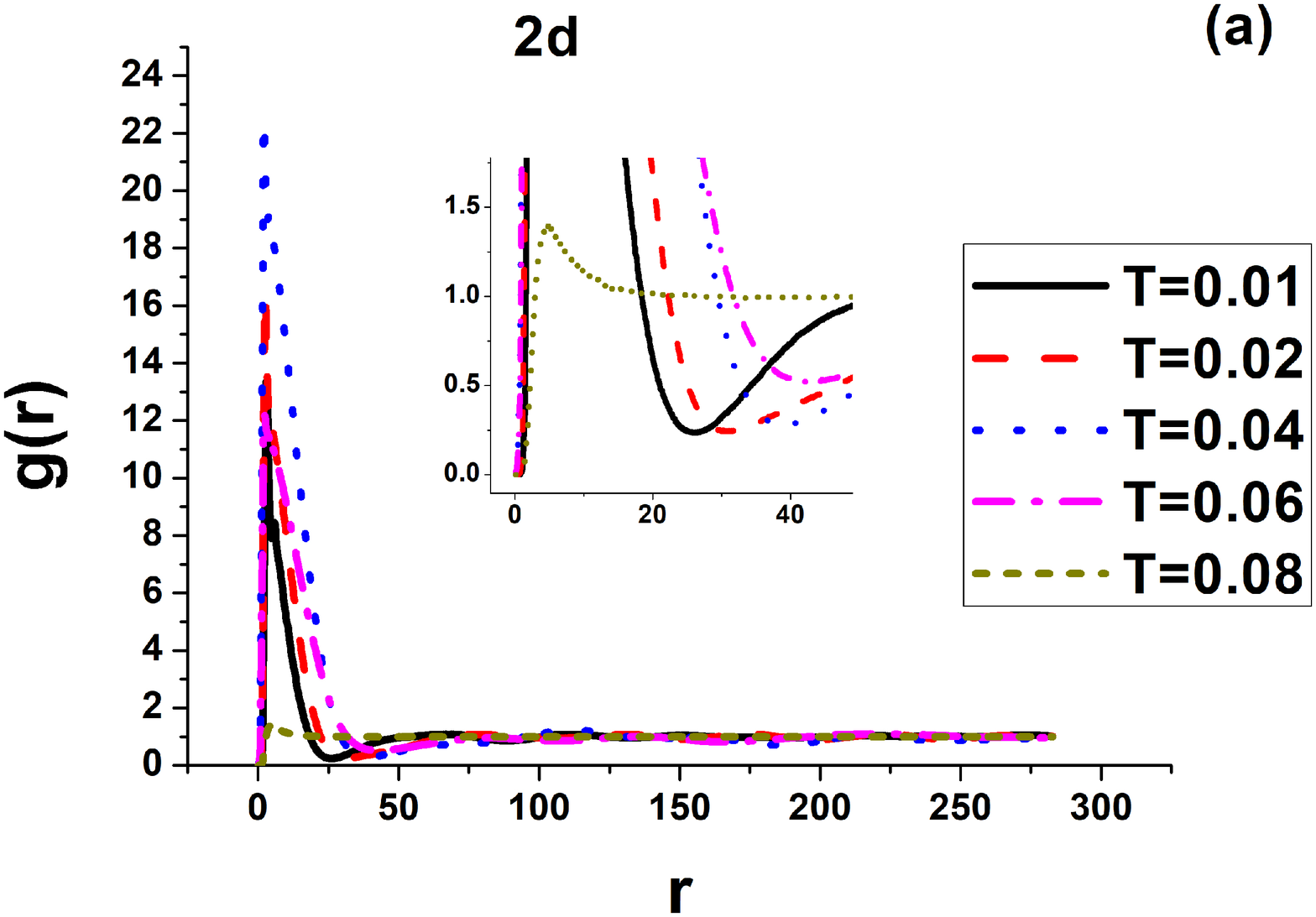}%

\includegraphics[width=8cm]{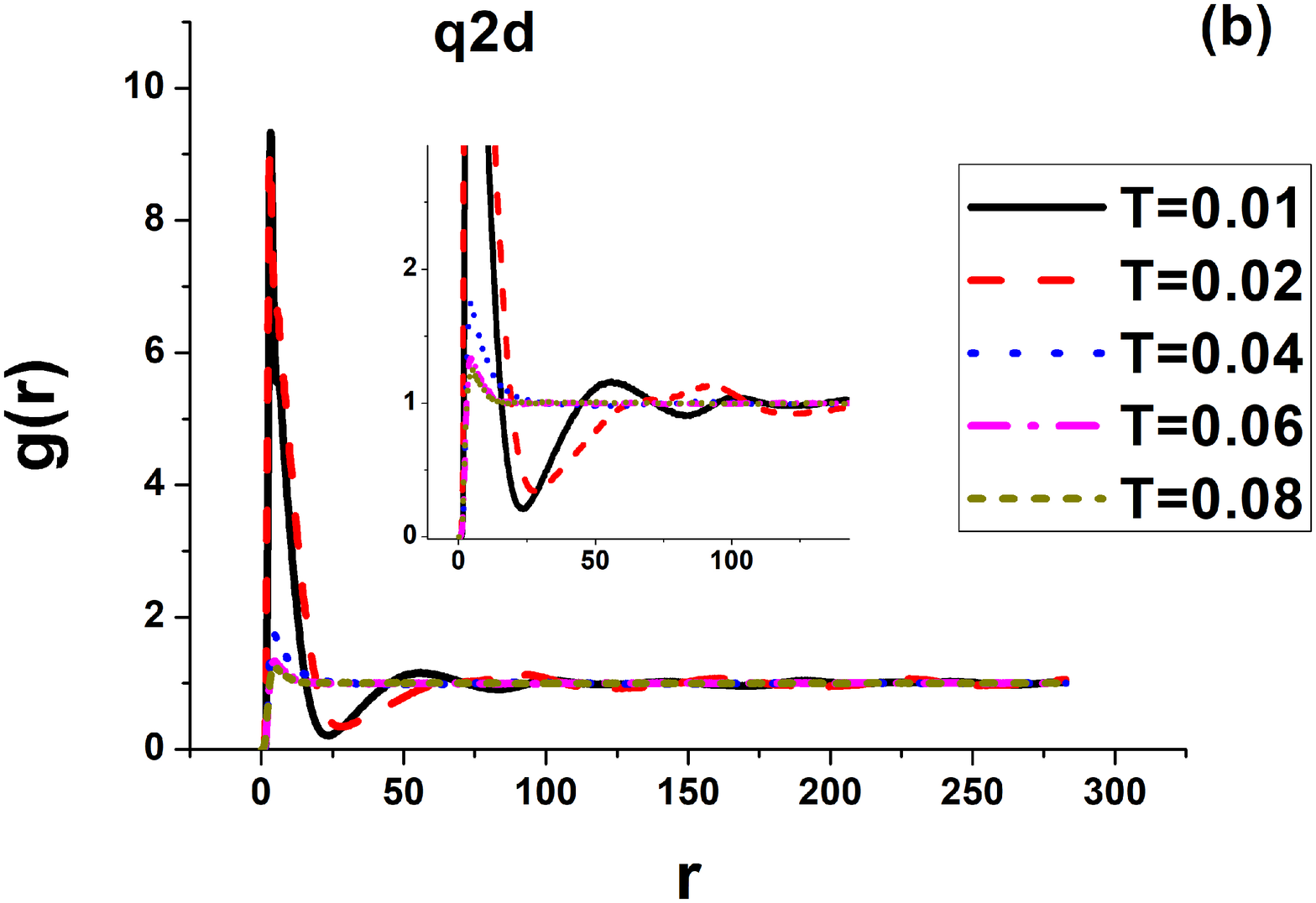}%

\includegraphics[width=8cm]{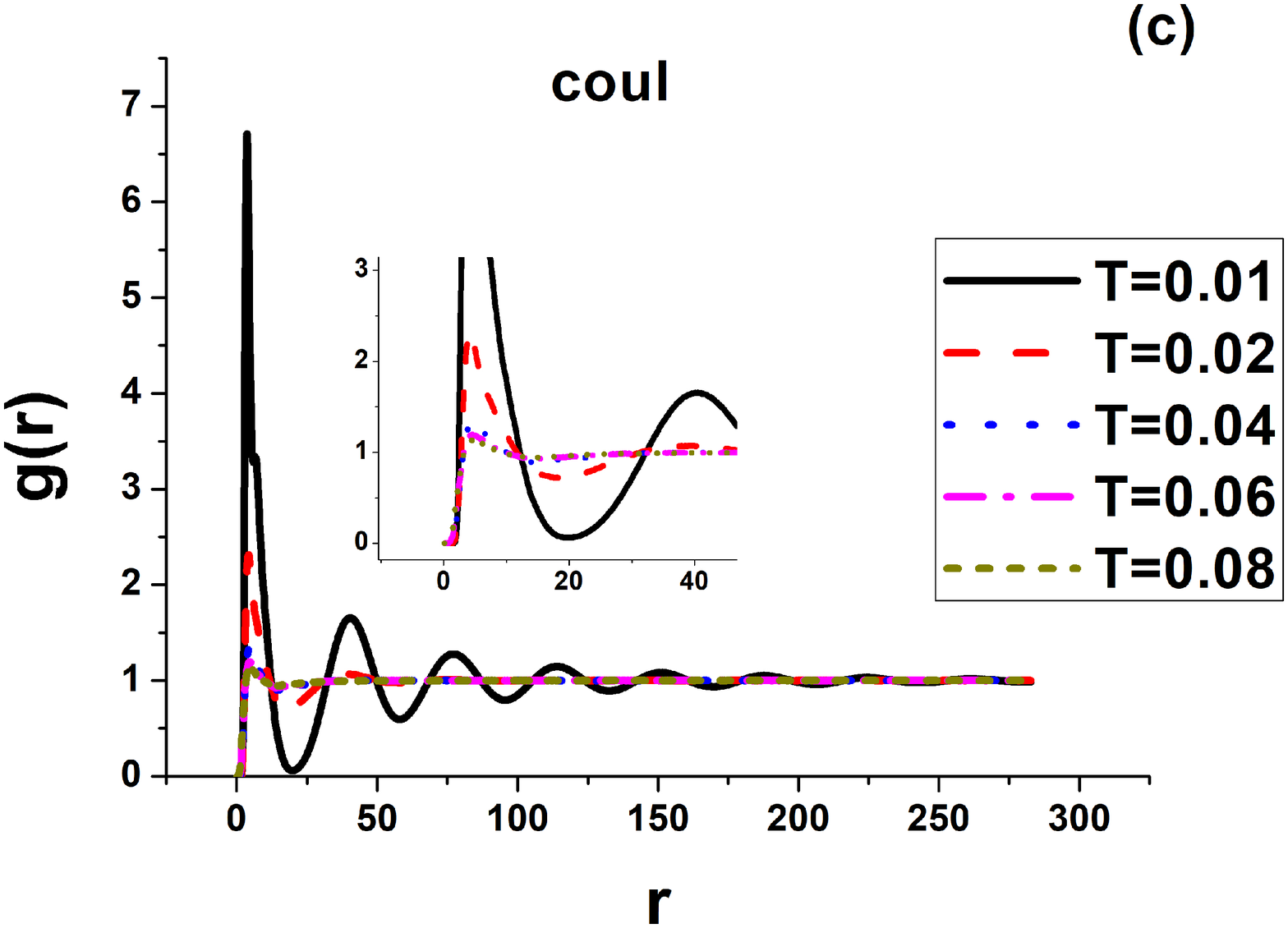}%

\caption{Temperature dependence of the RDF of (a) 2d, (b) q2d and
(c) coul systems at density $\rho=0.01$.}
\end{figure}

The same procedure was applied to the systems at density
$\rho=0.2$. The RDFs of 2d, q2d and coul systems at this density
are given in Fig. 14. In this case 2d and q2d consist of a single
cluster, while the coul system is in the triangular crystal phase.
One can see that the melting temperatures of the 2d and q2d
systems are extremely high compared with the coul one:
$T_{2d}=0.4$, $T_{q2d}=0.3$ and $T_{coul}=0.01$. Therefore, taking
into account the long range interactions in the system
dramatically changes not only its structure, but also
thermodynamical properties.

\begin{figure}\label{rdf-r02-t-2d}
\includegraphics[width=8cm]{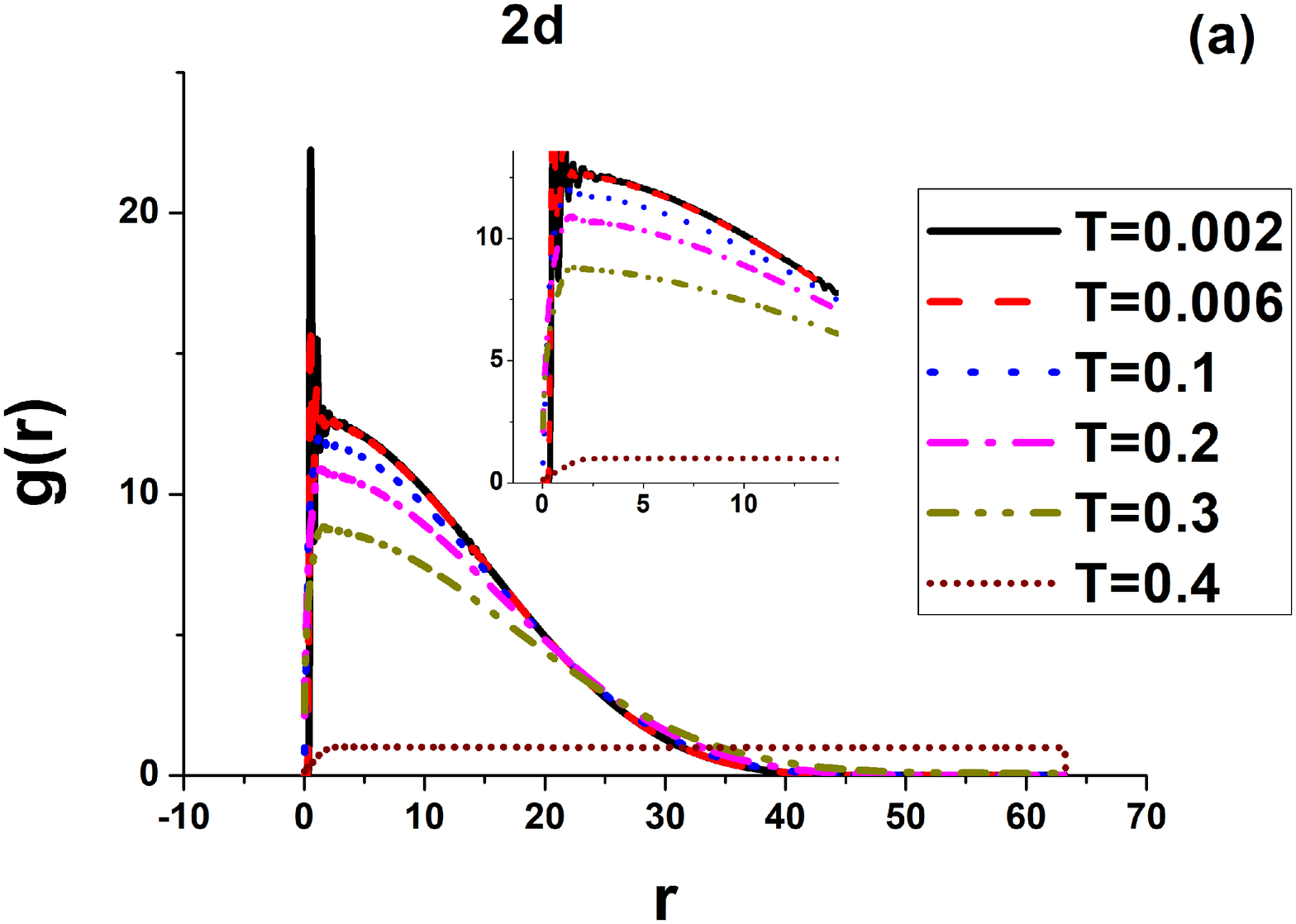}%

\includegraphics[width=8cm]{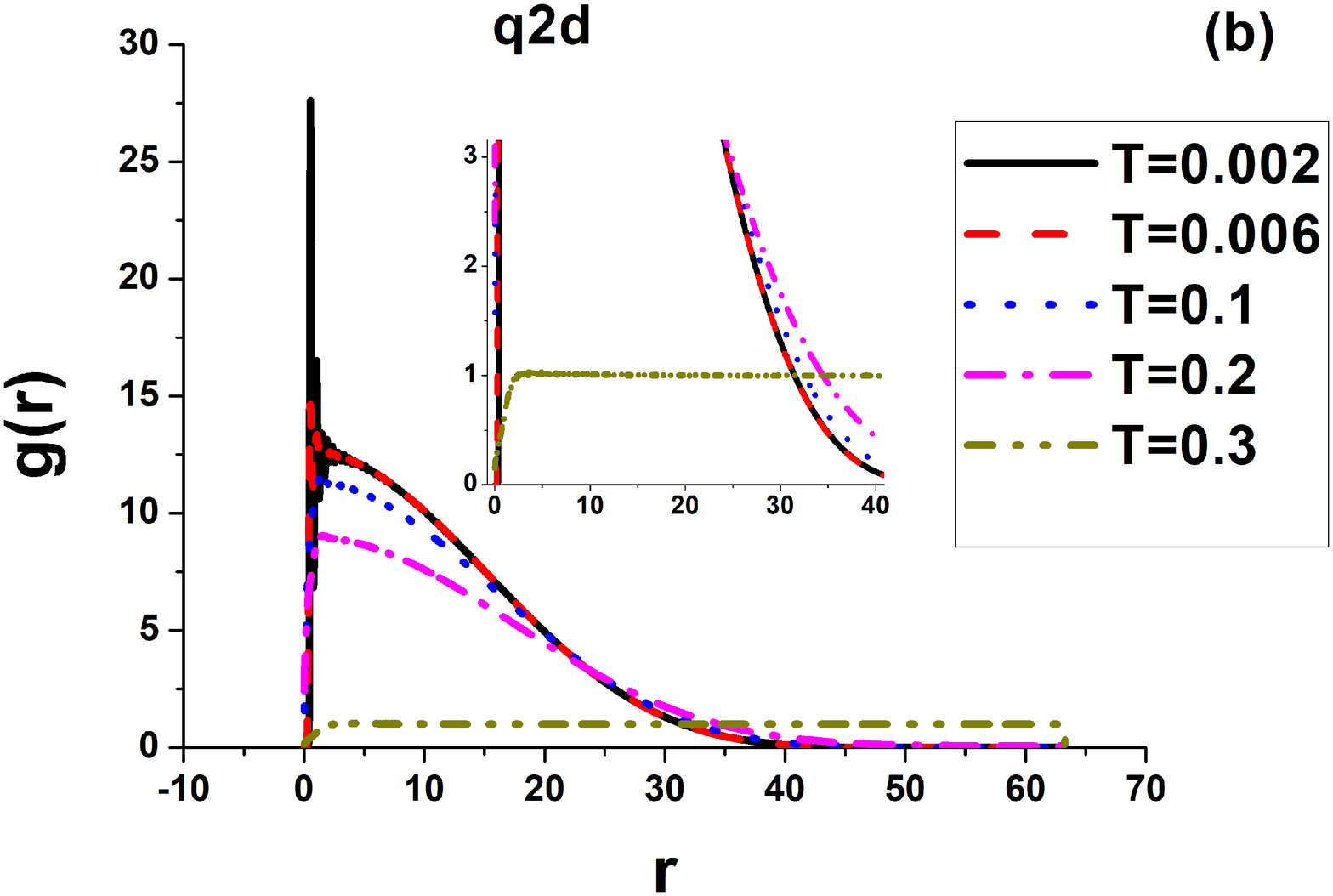}%

\includegraphics[width=8cm]{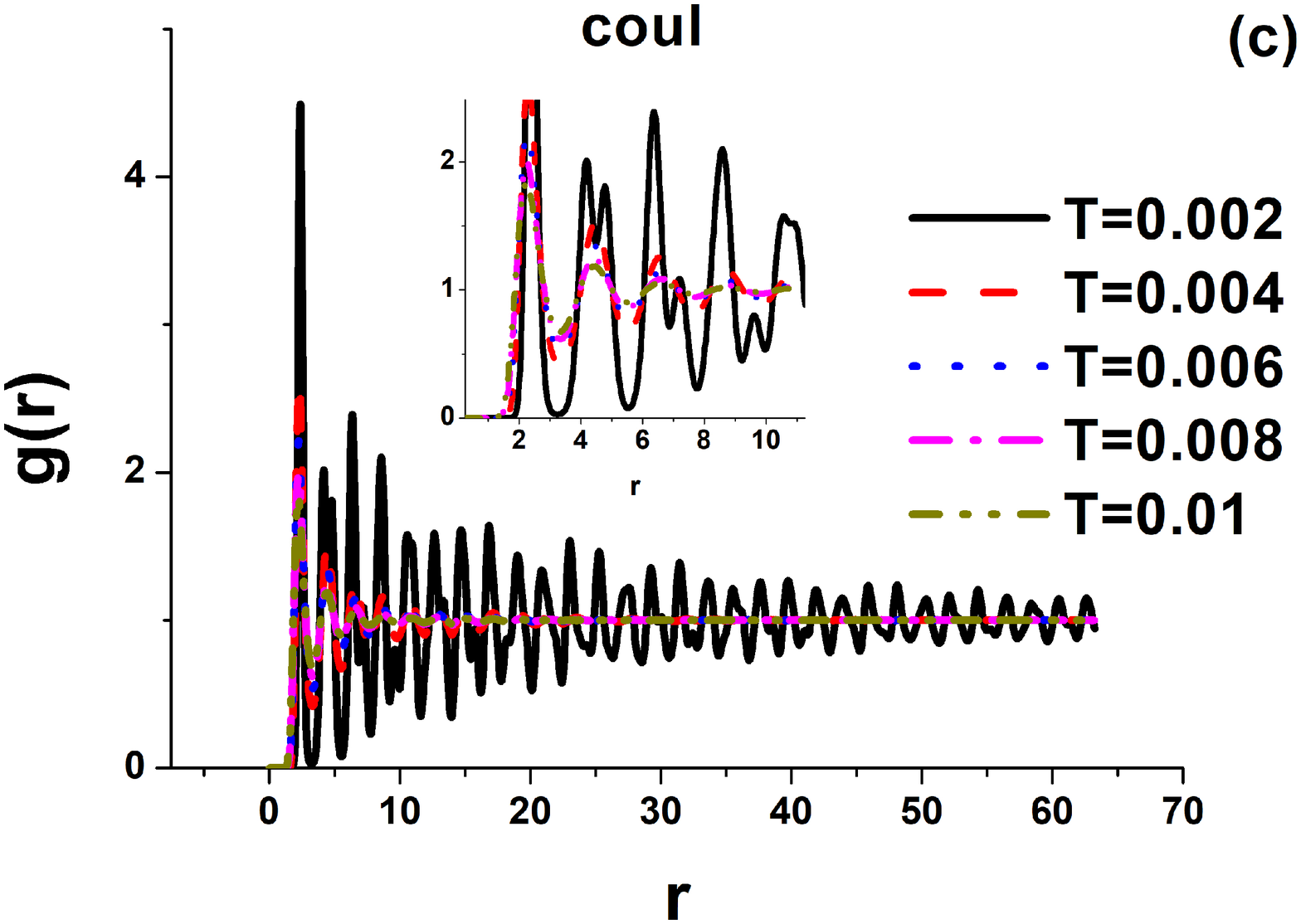}%

\caption{Temperature dependence of the RDF of (a) 2d, (b) q2d and
(c) coul systems at density $\rho=0.2$.}
\end{figure}

\section{Conclusions}

In the present paper we investigated a system of vortices in a
thin superconductive film interacting via  multiscale potential.
The following results were obtained.

1. In contrast with Refs. \cite{meng1,meng2,meng3} we took into
account  long-range interactions in the system (Eq. (2)). In order
to do this we introduced a coulombic term (by adding and
subtracting) into the interaction potential. In order to perform
calculations with model (2) two changes needed doing. Firstly we
had to substitute the 2d system with a q2d one, i.e. the system
became 3d, but the vortices were stuck to the plane by the LJ
walls. Secondly, to evaluate the long-range forces in the coul
system we employed the PPPM method \cite{bib}. We found that a
change from a 2d to a q2d system did not affect the result, while
the behavior of the coul system was entirely different from both
2d and q2d ones. Based on this we concluded that the changes in
the system behavior were induced by the long-range interactions.

2. We found that taking into account the long-range interaction in
the system led to a substantial change in its phase diagram. In
the 2d and q2d systems a phase of circular clusters existed with
densities ranging from the lowest up to $\rho=1.4$. A liquid phase
appeared at higher densities. In the coul system the phase diagram
was richer. The phase of circular clusters was observed at
densities $\rho=0.01$ and $\rho=0.05$. At $\rho=0.1$ a percolating
cluster was observed (Nazca phase). At $\rho=0.12$ a phase with
holes (a cheese or anti-clump phase) took place. At densities from
$\rho=0.14$ up to $\rho=1.0$ the triangular crystal was stable. At
higher densities the system was in a liquid state. Comparing the
radial distribution functions of 2d, q2d and coul systems we saw
that taking into account  long-range interactions led to a
substantial change in the characteristic length parameters in the
system: the distance to the first RDF peak was more than four
times greater than in the 2d and q2d systems.

3. We studied the finite size effects on the ribbon phase at
$\rho=1.2$ and $1.4$. We found that large systems formed the phase
of circular crystals while small ones - the ribbon phase. We
concluded that  ribbons were formed when system size was less than
the average size of a circular cluster.

4. We investigated the thermal stability of  2d, q2d and coul
systems at two densities $\rho=0.01$ and $\rho=0.2$. At the former
density all systems demonstrated a phase of circular clusters. At
the latter one the 2d and q2d systems were still in the phase of
circular clusters while the coul system formed a triangular
crystal. We found that upon heating, the internal structure of
clusters broke first. On further heating the clusters spread and
the system became isotropic. We also found that the melting
temperature of the 2d system at $\rho=0.2$ was 40 times higher
than the melting temperature of the coul system triangular phase
at the same density.

\bigskip

This work was carried out using computing resources of the federal
collective usage centre 'Complex for simulation and data
processing formega-science facilities' at NRC 'Kurchatov
Institute', http://ckp.nrcki.ru, and supercomputers at Joint
Supercomputer Center of the Russian Academy of Sciences (JSCC
RAS). The work was supported by the Russian Foundation of Basic
Research (Grants No 17-02-00320 and 18-02-00981).

\end{document}